\documentclass[10pt, a4paper, oneside, headinclude, footinclude]{scrartcl}

\usepackage[
nochapters, 
beramono, 
pdfspacing, 
dottedtoc 
]{classicthesis} 

\usepackage{arsclassica} 

\usepackage[T1]{fontenc} 

\usepackage[utf8]{inputenc} 

\usepackage{graphicx} 
\graphicspath{{Figures/}} 

\usepackage{enumitem} 

\usepackage{lipsum} 

\usepackage{subcaption} 

\usepackage{amsmath,amssymb,amsthm} 

\theoremstyle{definition} 

\theoremstyle{plain} 

\theoremstyle{remark} 

\hypersetup{
colorlinks=true, breaklinks=true, bookmarks=true,bookmarksnumbered,
urlcolor=webbrown, linkcolor=RoyalBlue, citecolor=webgreen, 
pdftitle={}, 
pdfauthor={\textcopyright}, 
pdfsubject={}, 
pdfkeywords={}, 
pdfcreator={pdfLaTeX}, 
pdfproducer={LaTeX with hyperref and ClassicThesis} 
} 

\usepackage[left = 4cm, right = 4cm, top = 2cm, bottom = 2.8cm]{geometry}
\usepackage{authblk}
\usepackage{bm}
\usepackage{centernot}
\usepackage{float}
\usepackage[sectionbib]{natbib}
\usepackage{pgfplots}
\usepackage{tabularx}

\usetikzlibrary{arrows}
\usetikzlibrary{calc}
\usetikzlibrary{fit}
\usetikzlibrary{positioning}
\usetikzlibrary{shapes}

\bibpunct{(}{)}{;}{a}{}{;}

\title{\spacedallcaps{\large A Comparison of Economic Agent-Based Model Calibration Methods}}

\author[1, 2]{\spacedlowsmallcaps{Donovan Platt} \thanks{Corresponding author, donovan.platt@maths.ox.ac.uk}}

\affil[1]{\small Mathematical Institute, University of Oxford}
\affil[2]{\small Institute for New Economic Thinking (INET) at the Oxford Martin School}

\date{}

\begin{document}


\clearscrheadfoot

\cfoot[\pagemark]{\pagemark}


\maketitle
\setcounter{tocdepth}{2}

\vspace{-1.5cm}

\section*{\centerline{Abstract}}

\begin{abstract}
\noindent Interest in agent-based models of financial markets and the wider economy has increased consistently over the last few decades, in no small part due to their ability to reproduce a number of empirically-observed stylised facts that are not easily recovered by more traditional modelling approaches. Nevertheless, the agent-based modelling paradigm faces mounting criticism, focused particularly on the rigour of current validation and calibration practices, most of which remain qualitative and stylised fact-driven. While the literature on quantitative and data-driven approaches has seen significant expansion in recent years, most studies have focused on the introduction of new calibration methods that are neither benchmarked against existing alternatives nor rigorously tested in terms of the quality of the estimates they produce. We therefore compare a number of prominent ABM calibration methods, both established and novel, through a series of computational experiments in an attempt to determine the respective strengths and weaknesses of each approach and the overall quality of the resultant parameter estimates. We find that Bayesian estimation, though less popular in the literature, consistently outperforms frequentist, objective function-based approaches and results in reasonable parameter estimates in many contexts. Despite this, we also find that agent-based model calibration techniques require further development in order to definitively calibrate large-scale models. We therefore make suggestions for future research.

\end{abstract}

\vspace{1cm}

\noindent \textbf{Keywords}: Agent-based modelling, Calibration, Simulated minimum distance, Bayesian estimation

\vspace{0.5cm}

\noindent \textbf{JEL Classification}: C13 $\cdot$ C52


\section{Introduction}

The modelling of economic systems presents a significant challenge -- the heterogeneity of their constituent agents, the complex nature of the interactions within them, and the non-linearity of their emergent dynamics makes them extremely difficult, if not impossible to represent using traditional methods. As a result, the use of empirically-inconsistent assumptions\footnote{This includes, but is not limited to assumptions of perfect rationality and the existence of representative agents. A more detailed discussion on general equilibrium theory and criticisms thereof is presented by \citet{Farmer_Geanakoplos_2009}.} and a disregard for the heterogeneity and non-linearity that characterise economic systems has for many decades been the dominant paradigm in economic modelling \citep{Farmer_Geanakoplos_2009, Farmer_Foley_2009, Fagiolo_Roventini_2017}.

Such criticisms of traditional economic models, or more specifically those derived from general equilibrium theory and its various extensions\footnote{Dynamic stochastic general equilibrium (DSGE) models are a prominent, contemporary example.}, have become increasingly prominent in the wake of the Great Recession of the late 2000s, where such models, which had long been thought to be trustworthy, would be shown to be inadequate, both in predicting the possibility of a financial crisis and in providing concrete solutions to resolve it \citep{Farmer_Foley_2009, Geanakoplos_et_al_2012, Fagiolo_Roventini_2017}. In this context, the need for substantial improvements to existing methodologies or the development of a viable alternative is clear.

Recent advances in computing power, along with successes in other domains such as ecology, have resulted in the emergence of a growing community arguing that agent-based models (ABMs), which simulate systems at the level of individual agents and the interactions between them\footnote{See \citet{Macal_North_2010} for a brief introduction to agent-based modelling.}, may provide a more principled approach to the modelling of the economy \citep{Farmer_Foley_2009, Fagiolo_Roventini_2017}. Indeed, recent decades have seen the emergence of a wide variety of economic and financial ABMs that largely dispense with the unrealistic assumptions that characterise traditional approaches in favour of more realistic alternatives rooted in empirically-observed behaviours \citep{Chen_2003, LeBaron_2006}.

This paradigm shift has ultimately resulted in a degree of success. In more detail, ABMs are well-known for their ability to replicate empirically-observed stylised facts, or qualitative properties that appear consistently in empirically-measured data, despite the fact that such properties are not readily recovered using traditional approaches \citep{LeBaron_2006, Barde_2016}. The literature regarding such stylised facts is very well-developed. An authoritative survey in the context of financial time series is presented by \citet{Cont_2001} and indicates that empirically-observed asset returns typically have a fat-tailed distribution, demonstrate no serial autocorrelation and present evidence of volatility clustering. In a more general economic context, the distribution of firm sizes is known to follow a Zipf distribution \citep{Axtell_2001} and the distribution of firm growth rates is typically fat-tailed \citep{Dosi_et_al_2017}. The preceding examples are by no means exhaustive and, in general, recent years have seen the emergence of ABMs that are increasingly ambitious in scope and capable of reproducing increasingly large sets of stylised facts\footnote{See, for example, the Eurace \citep{Cincotti_et_al_2010} and Schumpeter Meeting Keynes \citep{Dosi_et_al_2010} models.}.

Despite the aforementioned successes, ABMs face strong criticisms of their own, focused particularly on the inadequacy of current validation and calibration practices \citep{Grazzini_Richiardi_2015}. In the vast majority of studies, particularly those that introduce large-scale models, validation procedures are qualitative in nature and seldom venture beyond the demonstration of a candidate model's ability to reproduce a set of empirically-observed stylised facts \citep{Panayi_et_al_2013, Guerini_Moneta_2017}. Calibration in such investigations is equally rudimentary, and typically takes the form of manual parameter adjustments or ad-hoc processes that aim to select parameters that allow the model to reproduce the set of stylised facts considered during validation\footnote{The procedure employed by \citet{Jacob_Leal_et_al_2016} is a representative example.}. While such stylised fact-centric methodologies may seem reasonable at first glance, the very large number of models able to recover a similar number of stylised facts, the wide variety of behavioural rules employed in these models, and the difficulty experienced in attempting to identify the causal effects of many behavioural rules on emergent dynamics, renders robust model comparison an impossibility when using qualitative, stylised fact-centric methods \citep{LeBaron_2006, Barde_2016, Lamperti_et_al_2017}. This leads to what is often referred to as the "wilderness of bounded rationality" problem \citep{Farmer_Geanakoplos_2009, Barde_2016}.

In response to these criticisms, a small, but growing literature dealing with more sophisticated quantitative validation and calibration techniques has emerged \citep{Fagiolo_et_al_2017}. While significant progress has been made, particularly in the last three years, this literature still suffers from a number of key weaknesses. Firstly, it is overly compartmentalised. By this, we mean that most publications within this research area focus on the proposal of new methods and seldom compare the proposed techniques to other contemporary alternatives. This, combined with the fact that the theoretical properties\footnote{In most cases, this would refer to the bias and consistency of the associated estimators.} of many of these new techniques are not well understood \citep{Grazzini_et_al_2017}, leaves the modeller with a difficult choice between a large number of methods with no obvious reason to favour one approach over another. Secondly, severe computational limitations have resulted in most techniques only ever being applied to highly-simplified models\footnote{Perhaps the most notable of these is the \citet{Brock_Hommes_1998} model.} that are several decades old and no longer a good representation of the current state of economic agent-based modelling \citep{Lamperti_et_al_2017, Fagiolo_et_al_2017}. This leads to significant doubt regarding the applicability of current methods to the large-scale models that now dominate the literature.

We therefore aim to address the above using a principled, yet practical approach. Specifically, we compare a number of prominent ABM calibration methods, both established and novel, through a series of computational experiments involving the calibration of various candidate models in an attempt to determine the respective strengths and weaknesses of each approach. Thereafter, we apply the most promising of the considered calibration techniques to a large-scale ABM of the UK housing market in order to assess the extent to which the performance achieved in the context of simple models is maintained when confronting state of the art ABMs. Through these computational experiments, we obtain a broader and unified understanding of the current state of ABM calibration and make suggestions for future research.

\section{Literature Review} \label{Literature_Review}

As previously alluded to, there is a significant overlap between ABM validation and calibration literature. In most cases, calibration is concerned with selecting model parameters that result in dynamics that are as close as possible to those observed in a particular dataset, measured according to some criterion\footnote{These could include a set of stylised facts or some objective function.}. The same criterion may then also be used for validation purposes. Therefore, while we focus exclusively on the problem of ABM calibration, many of these discussions would also be applicable to ABM validation.

\tikzstyle{cal_type} = [rectangle, draw, fill = black!20, text width = 5em, text centered, rounded corners, minimum height = 4em, minimum width = 7em]
\tikzstyle{cal_frame_A} = [rectangle, draw, fill = none, text width = 5em, text centered, rounded corners, minimum height = 7em, minimum width = 28em] 
\tikzstyle{line} = [draw, -latex']

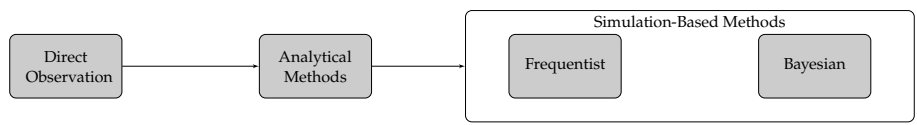
\begin{figure}[H]

\begin{center}

\scalebox{0.6}{

\begin{tikzpicture}[node distance = 3cm, auto] 

\node[cal_type] (Direct_Obs) {Direct \\ Observation};
\node[cal_type, right = of Direct_Obs] (Analytical) {Analytical Methods};
\node[cal_type, right = of Analytical] (Frequentist) {Frequentist};
\node[cal_type, right = of Frequentist] (Bayesian) {Bayesian};
\node[cal_frame_A, fit = (Frequentist)(Bayesian)] (Simulation) {\begin{minipage}[t][2cm]{4.2cm} Simulation-Based Methods \end{minipage}};

\path [line, align = left] (Direct_Obs) -- node [midway, below, scale=0.8] {} (Analytical);
\path [line, align = left] (Analytical) -- node [midway, below, scale=0.8] {} (Simulation);

\end{tikzpicture}

}

\end{center}

\caption{An illustration of the various calibration strategies one might consider when attempting to calibrate a particular ABM. \label{Calibration_Class_Overview}}

\end{figure}

At this point, it is worth noting that some authors use the terms calibration and estimation interchangeably, while others maintain that there are nuanced differences between them. As an example, \citet{Grazzini_Richiardi_2015} suggest that calibration is concerned only with obtaining agreement between the dynamics of a model and those of real-world data, while estimation additionally aims to ensure that the obtained model parameter values are an accurate reflection of the parameter values associated with the real-world data-generating process. Estimation would thus place additional emphasis on the uncertainty surrounding the obtained parameter values, while this may not be of much concern in the case of calibration. As stated by \citet{Hansen_Heckman_1996}, however, the distinction between these terms is often inconsistent and not entirely clear. We will therefore use the terms interchangeably, since the methods employed are likely to be very similar in either case.

We now proceed with a comprehensive review of the ABM calibration literature\footnote{The interested reader may also wish to refer to the surveys of \citet{Lux_Zwinkels_2017} and \citet{Fagiolo_et_al_2017}.}, beginning with the categorisation of ABM calibration strategies into three distinct classes, illustrated in Figure \ref{Calibration_Class_Overview}, which we then discuss in more detail.

\subsection{Direct Observation}

Since ABMs model systems by directly simulating the interactions of their microconstituents, it often arises that a number of model parameters reflect directly observable (or easily inferable) quantities, such as the net worth of firms or the distribution of ages among homeowners in a particular country. In this case, sophisticated estimation techniques are not required and the parameter values can be easily read directly from the data.

For some relatively simple models, such as the CATS model considered by \citet{Bianchi_et_al_2007}, values for almost all of the model's parameters can be determined in this way. While the same is not true for more sophisticated models, it still often arises that a large number of parameter values can be determined using a similar strategy, such as in the UK housing market model proposed by \citet{Baptista_et_al_2016}, where this process is referred to as micro-calibration.

\subsection{Analytical Methods}

Most ABMs include behavioural rules that require parameters that do not represent directly observable quantities. Therefore, even if appropriate values for many of a model's parameters can be determined through direct observation, it would still be necessary to apply statistical estimation techniques to a given dataset to select appropriate values for the remaining parameters. A logical first choice might be maximum likelihood estimation, or alternatively the method of moments, given that they are fairly general and well-understood methods.

In a very limited number of cases, maximum likelihood estimation has been applied successfully to ABMs \citep{Alfarano_et_al_2005, Alfarano_et_al_2006, Alfarano_et_al_2007}, but in general it and related methods are not appropriate as they rely on obtaining an analytical expression of the joint density function of time series simulated by the model. This is possible only for very simple models and even then is nontrivial.

\subsection{Simulation-Based Methods}

Owing to the fact that most ABMs of interest are incompatible with methods requiring analytical solutions for key model properties, it is inevitable that methods involving the generation of simulated data will need to be considered. Such approaches represent the key focus of most recent literature and we thus discuss them and related issues in extensive detail.

\subsubsection{Frequentist Inference: Traditional Approaches to Simulated Minimum Distance}

The vast majority of existing calibration attempts have adopted a frequentist approach, often employing variations of what are commonly referred to as simulated minimum distance (SMD) methods \citep{Grazzini_Richiardi_2015, Grazzini_et_al_2017}. Broadly speaking, these methods involve the construction of an objective function that measures the distance between simulated and measured time series for a given set of parameters, followed by the application of optimisation methods. More precisely, we have
\begin{equation}
\text{argmin}_{\bm{\theta} \in \bm{\Theta}} f(\bm{\theta}),
\end{equation}
where $\bm{\theta}$ is a vector of parameters, $\bm{\Theta}$ is the space of feasible parameters, and $f:\bm{\Theta} \rightarrow \mathbb{R}$ is a function measuring the distance between real and simulated time series.

While truly general and standardised methods are yet to appear in the context of economic ABMs, methods which consider weighted sums of the squared errors between simulated and empirically-measured moments (or other quantities that can be estimated from time series data) rose to prominence in early literature \citep{Gilli_Winker_2003} and the method of simulated moments (MSM) has been applied in numerous investigations\footnote{See \citet{Franke_2009}, \citet{Franke_Westerhoff_2012}, \citet{Fabretti_2013}, \citet{Grazzini_Richiardi_2015}, \citet{Chen_Lux_2016} and \citet{Platt_Gebbie_2018} for examples.}. Key motivations for the consideration of MSM include its prevalence throughout econometric literature, its transparency \citep{Franke_2009}, and its well-understood mathematical properties\footnote{The estimator is both consistent and asymptotically normal \citep{McFadden_1989}.}. 

MSM is not without a number of shortcomings, however. The single most significant concern is that the selection of moments is entirely arbitrary and a given set of moments will only represent a limited number of aggregate properties of the data and may therefore not sufficiently capture important dynamics. Further, the weight matrix obtained using the methodology described by \citet{Winker_et_al_2007}, the standard method in this context, may lead to numerical challenges in some cases due to the inversion of near-singular matrices \citep{Fabretti_2013, Platt_Gebbie_2018}.

A related approach is indirect inference (II), introduced by \citet{Gourieroux_et_al_1993}\footnote{An example of the application of II in the context of ABMs is presented by \citet{Bianchi_et_al_2007}.}. Rather than relying on estimated moments, II involves the use of what is called an auxiliary model, essentially a simple model that is amenable to estimation via analytical methods such as maximum likelihood. An objective function is constructed by estimating the auxiliary model on both empirically-measured and simulated data and comparing the obtained parameters, with minimisation implying the greatest similarity between the two sets of estimated parameters. In general, II faces similar criticisms to those levelled against MSM \citep{Grazzini_et_al_2017}, since it involves the selection of an arbitrary auxiliary model.

\subsubsection{Frequentist Inference: New Approaches to Simulated Minimum Distance}

As previously stated, traditional approaches to SMD such as MSM and II suffer from a number of weaknesses, the most significant of which is the need to select an arbitrary set of moments or auxiliary model. Not surprisingly, a number of alternatives have emerged aiming to address this particular problem, focusing on the structure of a given time series and its patterns rather than aggregate properties in an attempt to exploit the full informational content of the data.

Among the most straightforward of these alternatives is choosing the objective function to be the sum of appropriately-weighted squared differences between the values of the simulated and empirically-measured time series at a number of points in time, as has been done by \citet{Recchioni_et_al_2015}. More sophisticated approaches to quantifying differences in the structure of simulated and empirically-measured time series are presented by \citet{Lamperti_2017} and \citet{Barde_2017} in the form of information-theoretic criteria called the generalised subtracted L-divergence (GSL-div) and Markov information criterion (MIC) respectively\footnote{More detailed discussions relating to the GSL-div and MIC can be found in Section \ref{Implemented_Methods}.}. Recent years have also seen the emergence of attempts at comparing the causal mechanisms underlying real and simulated data through the use of SVAR regressions, as suggested by \citet{Guerini_Moneta_2017}.

While the above approaches largely succeed in attenuating concerns related to the selection of arbitrary moments or auxiliary models, they introduce new challenges. While MSM and II have well-understood theoretical properties, many of the recently introduced alternatives have not been subjected to rigorous mathematical analyses \citep{Grazzini_et_al_2017}. As a consequence, comparisons between different calibration methods tend to be non-trivial, as many of the aforementioned techniques have few conceptual similarities, and most arguments in favour of a particular method are likely to be superficial. This leaves the modeller with a choice between a large number of potential calibration approaches, with very little evidence available on which to base such a decision.

\subsubsection{Bayesian Inference}

Although less common in the literature, an additional alternative to traditional SMD methods is Bayesian inference, which does not require the selection of arbitrary aggregate features and also allows for the incorporation of known, prior information regarding the parameter values \citep{Grazzini_et_al_2017}. This is achieved through the following application of Bayes' theorem:
\begin{equation}
\mathbb{P}(\bm{\theta} | \bm{X}) \propto \mathbb{P}(\bm{X} | \bm{\theta})\mathbb{P}(\bm{\theta}),
\end{equation}
where $\bm{X}$ represents empirically-measured data, $\mathbb{P}(\bm{\theta})$ represents one's prior views regarding the parameter values, and $\mathbb{P}(\bm{X} | \bm{\theta})$ is the likelihood of the model generating $\bm{X}$ when initialised using parameter set $\bm{\theta}$. In contrast to SMD methods, which produce point estimates, it should be apparent that we do not obtain a single parameter set, but rather a distribution. This distribution can, however, be analysed to produce a suitable point estimate by taking the mean, median or mode. 

The most significant challenge presented by the approach is the estimation of the likelihood\footnote{It is worth noting that it is also possible to maximise $\mathbb{P}(\bm{X} | \bm{\theta})$ directly, rather than employing it within a Bayesian framework, resulting in a simulation-based approximation to maximum likelihood estimation. While \citet{Kukacka_Barunik_2017} apply such a procedure to the \citet{Brock_Hommes_1998} model, achieving a degree of success, Bayesian methods are, in general, more robust to overfitting and thus better suited to more complex models \citep{Murphy_2012}.}, $\mathbb{P}(\bm{X} | \bm{\theta})$. \citet{Grazzini_et_al_2017} indicate that non-parametric methods such as kernel density estimation (KDE) may be too computationally demanding to be feasible in most ABM applications, necessitating fairly strong assumptions\footnote{We describe the approach of \citet{Grazzini_et_al_2017} in more detail in Section \ref{Implemented_Methods}.}, while parametric approaches are often not flexible enough to accurately represent the distribution.

\subsubsection{Addressing Computational Difficulties}

While the lack of comparisons between newer methods is indeed a significant weakness of the ABM calibration literature, computational difficulties remain an even larger obstacle in the development of robust and widely-applicable ABM calibration strategies. Since most models of interest are likely to be costly to simulate and given the large amount of data that needs to be generated for comparison in most ABM calibration frameworks, few attempts have been made to calibrate large-scale ABMs. Instead, most investigations favour proof-of-concept demonstrations on simpler, closed-form models, such as those of \citet{Brock_Hommes_1998} and \citet{Farmer_Joshi_2002}. Ultimately, despite the conceptual similarities between large-scale models and simpler variants, the extent to which existing results can be generalised remains an open question.

Recently, there has been a relatively modest increase in the awareness of this point within some groups and increasing emphasis has been placed on addressing these computational difficulties in some circles \citep{Grazzini_et_al_2017, Lamperti_et_al_2017}. A somewhat promising suggestion is the use of surrogate modelling to circumvent the intensive ABM simulation process, with representative examples including Gaussian process interpolation, also known as kriging \citep{Salle_Yildizoglu_2014}\footnote{Refer to \citet{Barde_VanDerHoog_2017} for a first attempt at using similar methods to calibrate a large-scale economic ABM.}, and the more general machine learning surrogate approach of \citet{Lamperti_et_al_2017}.

Recent years have also seen the emergence of cloud computing platforms, such as Amazon Web Services, which provide users with access to computing resources that are essentially rented for a required task, rather than purchased outright, granting access to computing power several orders of magnitude greater than may be possible through traditional on-site means. Therefore, the use of such services may provide researchers with the means to tackle more ambitious calibration problems without the need to resort to surrogate modelling. This is in fact the approach that has been taken in this investigation.

\section{Experimental Procedure} \label{Experimental_Procedure}

In order to effectively compare various calibration techniques, it is necessary to develop a series of tests that quantify the notion of calibration success and result in measures that can be directly compared. One might assume that a natural approach would be the calibration of a set of candidate models to empirically-observed data using a variety of different approaches and assessing the resulting goodness of fit. Unfortunately, such an approach is likely to be suboptimal for a number of reasons. 

Firstly, regardless of the quality and sophistication of a candidate model, it is likely to be misspecified to some extent when compared to the true data-generating process, especially in an economic context. As an example, a given model may be overly-simplified and fail to capture the nuances of the underlying data-generating process, resulting in a poor fit to the data regardless of the merits of each calibration method. Secondly, the notion of goodness of fit is difficult to quantify in the context of ABMs. Indeed, every SMD objective function is an attempt at quantifying it, with each differing in what they consider to be the most important characteristics of the data. We therefore introduce a general approach to calibration method comparison that addresses the above concerns.

\subsection{Loss Function Construction and Comparison Procedure}

We begin by letting $\bm{X}^{s}_{i}(\bm{\theta}, T)$ be the output of a candidate model\footnote{In this case, we consider univariate time series, but analogous arguments apply to the case of panel data.}, $M$, for parameter set $\bm{\theta}$, output length $T$, and random seed $i$, where the use of a superscript $s$ indicates that the quantity being described is derived from or related to simulated rather than real data. Since empirically-observed data is nothing more than a single realisation of the true data-generating process, which may itself be viewed as a model with its own set of parameters, it follows that we may consider $\bm{X} = \bm{X}^{s}_{i^{*}}(\bm{\theta}_{true}, T_{emp})$ as a proxy for real data to which $M$ may be calibrated.

In this case, we are essentially calibrating a perfectly specified model to data for which the true set of parameters, $\bm{\theta}_{true}$, is known. It can be argued that a good calibration method would, in this idealised setting, be able to recover the true set of parameters to some extent and that methods which produce estimates closer to $\bm{\theta}_{true}$ would be considered superior. It also follows that methods which perform well in this context are far more likely to perform well in the more realistic case of a misspecified model. We therefore define the loss function
\begin{equation}
L(\bm{\theta}_{true}, \hat{\bm{\theta}}) = || \bm{\theta}_{true} - \hat{\bm{\theta}} ||_{2},
\end{equation}
where $\hat{\bm{\theta}}$ is the parameter estimate produced by a given calibration method. Therefore, using this approach, we not only address concerns related to misspecified models, but are also able to directly compare calibration methods according to their associated loss function values.

Of course, we are not suggesting that this is a definitive tool for calibration method comparison, but do believe it to be a principled approach that is able to provide meaningful insights in this context.

Given that we have now defined a metric that allows calibration methods to be directly compared for a given model, $M$, and true parameter set, $\bm{\theta}_{true}$, it is relatively straightforward to develop a comprehensive series of comparative tests. 

We begin by noting that the difficulty of a given test depends on three factors: the complexity of the dynamics produced by $M$, the number of free parameters in $\bm{\theta}_{true}$, and the length of the time series data to which $M$ is calibrated, $T_{emp}$. We must therefore consider a set of models $\bm{M}$, where each model differs in the complexity and overall nature of its resultant dynamics, and a variety of true parameter sets of different cardinalities for each model\footnote{Note that we do not vary the empirical time series length in our experiments, since the effects associated with the number of free parameters and complexity of the model dynamics are generally dominant.}. We can then determine the loss function values associated with each calibration method for the various models and true parameter sets. This will ultimately provide insight into which calibration techniques deliver the best performance and in which situations.

At this point, it should be noted that the models we consider will, in general, not be ergodic\footnote{The interested reader should refer to \citet{Grazzini_2012} for a detailed discussion on ergodicity and its relationship with ABM calibration.}. Therefore, each calibration experiment involves the comparison of our proxy for real data, $\bm{X}^{s}_{i^{*}}(\bm{\theta}_{true}, T_{emp})$, with an ensemble of $R$ Monte Carlo replications for each candidate set of parameters, $\bm{X}^{s}_{i}(\bm{\theta}, T_{sim}), i = i_{0}, i_{0} + 1, \dots, i_{0} + R - 1$, where $i_{0} \in \mathbb{N}$ and we assume that $i^{*} \centernot\in \{i_{0}, i_{0} + 1, \dots, i_{0} + R - 1\}$.

\subsection{Implemented Models \label{Implemented_Models}}

We now provide descriptions of the implemented models and brief motivations for their consideration in this study. The first four are computationally inexpensive, appear frequently in the existing calibration literature, and are capable of producing dynamics ranging from very basic to relatively sophisticated. This initial set is selected to identify the most promising among the considered methods and those that perform well in this simplified context will then be applied to a far more computationally expensive, large-scale ABM of the UK housing market.

The computationally inexpensive nature of the initial set of models allows us to perform a thorough series of tests involving all of the considered calibration methods in a realistic period of time, after which only methods that are likely to be successful are applied in the more complex setting. Indeed, methods which do not perform well in this simplified context will almost certainly perform poorly when applied to the UK housing market model and we therefore need not waste computational resources on such tests.

\subsubsection{AR$(1)$ Model}

The first model we consider is an autoregressive model of order $1$, given by
\begin{equation}
x_{t + 1}= a_1 x_{t} + \epsilon_{t + 1},
\end{equation}
where $\epsilon_t \sim \mathcal{N}(0, 1)$. 

It should be apparent that the above is a basic, single parameter model capable of producing a limited range of dynamics and that its calibration should be a trivial exercise. It has been included in order to create a baseline test for which we expect all of the considered calibration methods to perform well.

\subsubsection{ARMA$(2, 2)$-ARCH$(2)$ Model}

The second model we consider is an ARMA$(2, 2)$ model with ARCH$(2)$ errors, given by
\begin{equation}
\setlength{\jot}{10pt}
\begin{split}
x_{t + 1} &= a_0 + a_1 x_{t} + a_2 x_{t - 1} + b_1 \sigma_{t} \epsilon_{t} + b_2 \sigma_{t - 1} \epsilon_{t - 1} + \sigma_{t + 1} \epsilon_{t + 1}, \\
\sigma_{t + 1} ^2 &= c_0 + c_1 \epsilon_t ^2 + c_2 \epsilon_{t - 1} ^2, \\
\end{split}
\end{equation}
where $\epsilon_t \sim \mathcal{N}(0, 1)$. 

The above is a logical successor to the previously considered AR$(1)$ model, as it is drawn from the same general class, traditional econometric time series models, but is capable of producing more nuanced dynamics and has a parameter space of significantly larger cardinality. Additionally, the calibration of such models is relatively straightforward using analytical methods. It would thus be worthwhile to assess the performance of simulation-based methods in the context of models known to be amenable to accurate estimation using analytical approaches.

A similar model was considered by \citet{Barde_2017} when testing the MIC, though it was not used in full calibration experiments and involved the comparison of the obtained objective function values for a limited number of parameter combinations.

\subsubsection{Random Walks with Structural Breaks}

The third model we consider is a random walk capable of replicating simple structural breaks, given by
\begin{equation}
x_{t + 1}= x_{t} + d_{t + 1} + \epsilon_{t + 1},
\end{equation}
where $\epsilon_t \sim \mathcal{N}(0, \sigma_t ^2)$ and
\begin{equation}
d_t, \sigma_t = 
\begin{cases}
d_1, \sigma_1 & t \leq \tau \\
d_2, \sigma_2 & t > \tau.
\end{cases}
\end{equation}

Despite the simplicity of the model's equations, its replication of structural breaks and non-stationarity should present a significant challenge to the considered methods. 

It should also be noted that the above model was used by \citet{Lamperti_2017} to test the GSL-div, though, in much the same way that the ARMA$(2, 2)$-ARCH$(2)$ model was employed by \citet{Barde_2017} during the testing of the MIC, this involves the comparison of the objective function values obtained for a limited number of parameter value combinations rather than attempts at calibration.

\subsubsection{\citet{Brock_Hommes_1998} Model}

The fourth model we consider is the heterogenous agent model proposed by \citet{Brock_Hommes_1998}, which has a closed-form solution given by
\begin{equation}
\setlength{\jot}{10pt}
\begin{split}
x_{t + 1} &= \frac{1}{1 + r} \sum_{h = 1} ^{H} n_{h, t + 1} (g_h x_t + b_h) + \epsilon_{t + 1}, \\
n_{h, t + 1} &= \frac{\exp(\beta U_{h, t})}{\sum_{h = 1} ^{H}\exp(\beta U_{h, t})}, \\
U_{h, t} &= (x_t - R x_{t - 1})(g_h x_{t - 2} + b_h - R x_{t - 1}), \\
\end{split}
\end{equation}
where $\epsilon_t \sim \mathcal{N}(0, \sigma^2)$ and $R = 1 + r$.

The above is well-known in the literature as an early example of a class of ABMs that attempt to model the trading of assets on an artificial stock market by simulating the interactions of heterogenous traders that follow various trading strategies\footnote{The interested reader should refer to \citet{Brock_Hommes_1998} for a detailed discussion of the model's underlying assumptions and the derivation of the above closed-form solution.}. 

Each strategy, $h$, has an associated trend following component, $g_h$, and bias, $b_h$, both of which are real-valued parameters that are of particular interest in our investigation. The model also includes positive-valued parameters that affect all trader agents, regardless of the strategy they are currently employing, specifically $\beta$, which controls the rate at which agents switch between various strategies, and the prevailing market interest rate, $r$. 

Despite its relative simplicity, the \citet{Brock_Hommes_1998} model is capable of producing a range of sophisticated dynamics, including chaotic behaviours, and is computationally inexpensive to simulate, unlike most contemporary ABMs. These desirable features have led to it being a popular choice in many ABM calibration exercises throughout the literature, making it a natural inclusion in this investigation.

\subsubsection{INET Oxford Housing Market Model}

The fifth and final model we consider is a large-scale\footnote{The model has roughly $100$ parameters.} ABM of the UK housing market, which was introduced in the working paper by \citet{Baptista_et_al_2016}. As previously discussed, the consideration of such models in the calibration literature is still very rare, with most recent attempts focusing on far simpler alternatives, such as the \citet{Brock_Hommes_1998} model. Since the estimation of large-scale models remains an open problem and is the ultimate goal of current research in the field, attempts to calibrate this model will form a central component of this investigation.

Unfortunately, the sophistication of the model is such that we are unable to describe it here. Nevertheless, the interested reader should refer to the original working paper and the publicly available source code, which can be found at: \url{https://github.com/EconomicSL/housing-model}.

\subsection{Implemented Methods} \label{Implemented_Methods}

Given that we have now described the overall experimental procedure and selected models in detail, we finally discuss each of the implemented calibration methods, which have been selected such that each major area of the literature is represented.

\subsubsection{SMD Objective Functions}

Since SMD is the most popular ABM calibration paradigm, it is appropriate that it features prominently in this investigation. We therefore consider several SMD objective functions, including:
\begin{itemize}
\item MSM. Despite its need for the selection of an arbitrary set of moments, MSM remains a popular choice throughout the ABM calibration and econometric literature in general. Its inclusion is therefore essential as a benchmark against which to compare newer approaches. While there are a number of different interpretations of the MSM framework and many moment sets that could potentially be considered, we ultimately chose the implementation described by \citet{Franke_2009}, combined with the moment set proposed by \citet{Chen_Lux_2016}, consisting of the variance, kurtosis, autocorrelation coefficients for the raw series, absolute value series and squared series at lag $1$, and the autocorrelation coefficients for the absolute value series and squared series at lag $5$.
\item The GSL-div \citep{Lamperti_2017}. In essence, this recently introduced information theoretic criterion compares the distributions of temporal patterns occurring in different time series. It achieves this by discretising both the simulated and empirically-observed data into series of windows of a given length, viewing each window as a word in a corresponding vocabulary, and proceeding to calculate the subtracted L-divergence between the word distributions implied by each dataset using frequency-based estimators. Thereafter, these L-divergence estimates are weighted and summed for a desired number of window lengths.
\item The MIC \citep{Barde_2017}. Despite also being an information theoretic criterion and therefore direct competitor to the GSL-div, the MIC adopts a fundamentally different approach to achieve its goals and therefore warrants consideration in our investigation. In more detail, it constructs an $N$-th order Markov process based on the model-simulated data using an algorithm known as context tree weighting \citep{Willems_et_al_1995} and makes use of the obtained transition probabilities and the sum of the binary log scores along the length of the empirically-observed data to produce an estimate of the cross entropy.

\end{itemize}

\subsubsection{Optimisation Algorithms}

In the preceding subsection, we briefly discussed three distinct approaches that may be considered when constructing an objective function in the SMD framework. A second and equally important concern is the method used to minimise a given objective function. Unfortunately, the resulting optimisation problem is, in general, incredibly difficult. 

Firstly, the simulation-based nature of the aforementioned SMD objective functions means that we cannot make use of gradient-based methods or related approaches that require analytical expressions for the value of the objective function at a given point. This will ultimately force us to consider heuristic methods, which often produce solutions with properties that are not well-understood and do not guarantee convergence to a global minimum \citep{Gilli_Winker_2003, Fabretti_2013}. Secondly, and even more importantly, such methods typically require a significant number of objective function evaluations, each of which are very computationally expensive in our case.

Given that there are no best practices in the literature, we will consider (and ultimately compare) two contemporary heuristics, namely:
\begin{itemize}
\item Particle swarm optimisation, an evolutionary algorithm which mimics the flocking and swarming behaviours of organisms in ecological systems. Algorithms of this nature are now standard in the optimisation literature, with an accessible overview provided by \citet{Kaveh_2017}.
\item The approach of \citet{Knysh_Korkolis_2016}, a surrogate model method based on the work of \citet{Regis_Shoemaker_2005} and designed for the optimisation of expensive blackbox functions. The approach begins by efficiently sampling the parameter space using latin hypercube sampling and evaluating the objective function at the sampled points. Thereafter, it proceeds by constructing a computationally inexpensive approximation to the objective function using radial basis functions, which is then minimised to obtain an initial solution\footnote{Note that this is similar to the kriging approach of \citet{Salle_Yildizoglu_2014}, discussed in Section \ref{Literature_Review}.}. This solution is then gradually improved through the consideration of further evaluations of the original objective function in an iterative procedure known as the CORS algorithm.
\end{itemize}

\subsubsection{Bayesian Estimation}

As previously stated, the consideration of Bayesian inference is still relatively rare in the ABM calibration literature. Nevertheless, the work of \citet{Grazzini_et_al_2017} provides a first attempt at its use within this context and is thus worth comparing to the dominant SMD paradigm. 

In essence, the approach of \citet{Grazzini_et_al_2017} assumes that observations in each of the considered time series (model-simulated or empirically-observed) are i.i.d. random samples from unconditional distributions that characterise observations in each series. Under this assumption, KDE is used to construct a density function that characterises the distribution of observations in the simulated series for a given set of parameter values, which can then be used to determine the likelihood of the empirically-observed series for this parameter set. Applying Bayes theorem results in a posterior distribution that is amenable to sampling using a random walk Metropolis-Hastings algorithm.

While \citet{Grazzini_et_al_2017} consider only full posterior distributions, we require point estimates to ensure compatibility with our loss function. In most cases and particularly for parameters taking on continuous values, the mean or median are preferred, since the mode may be an atypical point \citep{Murphy_2012}. In our case, we consider the posterior mean, since it corresponds to the optimal estimate for the squared error loss and thus our chosen loss function.

\section{Results and Discussion}

In Section \ref{Experimental_Procedure}, we described a procedure for comparing the effectiveness of various ABM calibration techniques using computational experiments. We now present the results of these experiments and discuss their implications.

\subsection{Simple Time Series Models}

The first series of tests involves the application of all of the implemented calibration techniques to a set of simple time series models. While the main elements of the experimental procedure have been outlined in Section \ref{Experimental_Procedure}, we have also provided a more detailed overview of the technical details of these experiments in Appendix \ref{Experiment_Details}, including discussions related to dataset construction and the setting of the hyperparameters for each calibration method.

\subsubsection{AR$(1)$ Model}

We now proceed with the presentation of the results of our proposed comparative experiments, beginning with those obtained when attempting to calibrate the AR$(1)$ model. Since the model has only a single parameter, $a_1 \in [0, 1]$, and forms part of a baseline test that all of the considered methods are expected to perform well in, our discussion will be relatively brief.

\begin{table}[h]

\caption{Calibration Results for the AR$(1)$ Model} \label{AR_1_Results_Table}

\centering

\begin{tabularx}{\textwidth}{XXX}
\hline
& $a_1$ & $L(\hat{\bm{\theta}}, \bm{\theta}_{true})$\\
\hline
$\bm{\theta}_{true}$ & $0.7$ & $0$ \\
GSL-div/PS & $0.7618$ & $0.0618$ \\
GSL-div/KK & $0.7877$ & $0.0877$ \\
MSM/PS & $0.6568$ & $0.0432$ \\
MSM/KK & $0.6592$ & $0.0408$ \\
MIC/PS & $0.6599$ & $0.0401$ \\
MIC/KK & $0.6611$ & $0.0389$ \\
BE & $0.6672$ & $0.0328$ \\
\hline
\end{tabularx}

\end{table}

\begin{figure}[H]

\centering

\begin{subfigure}{1\textwidth}
	\centering
	\includegraphics[width=1\linewidth]{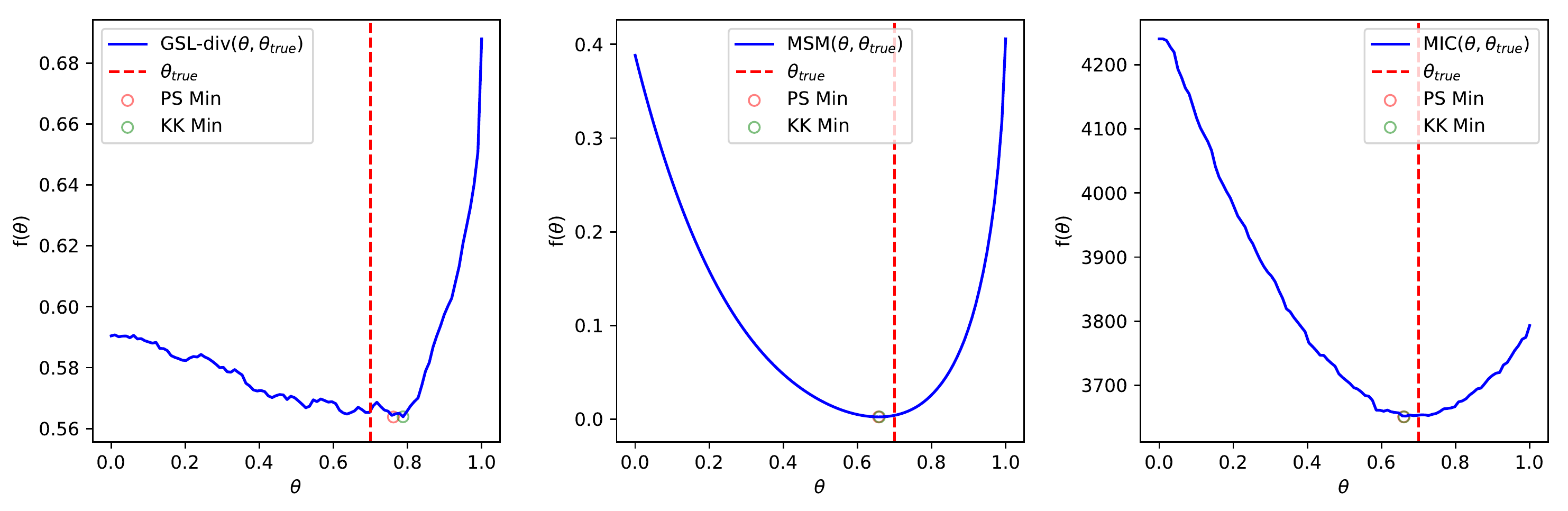}
	\caption{SMD objective function curves for parameter $a_1$.} \label{AR_1_Curves}
\end{subfigure}

\vspace{0.45cm}

\begin{subfigure}{1\textwidth}
	\centering
	\includegraphics[width=0.32\linewidth]{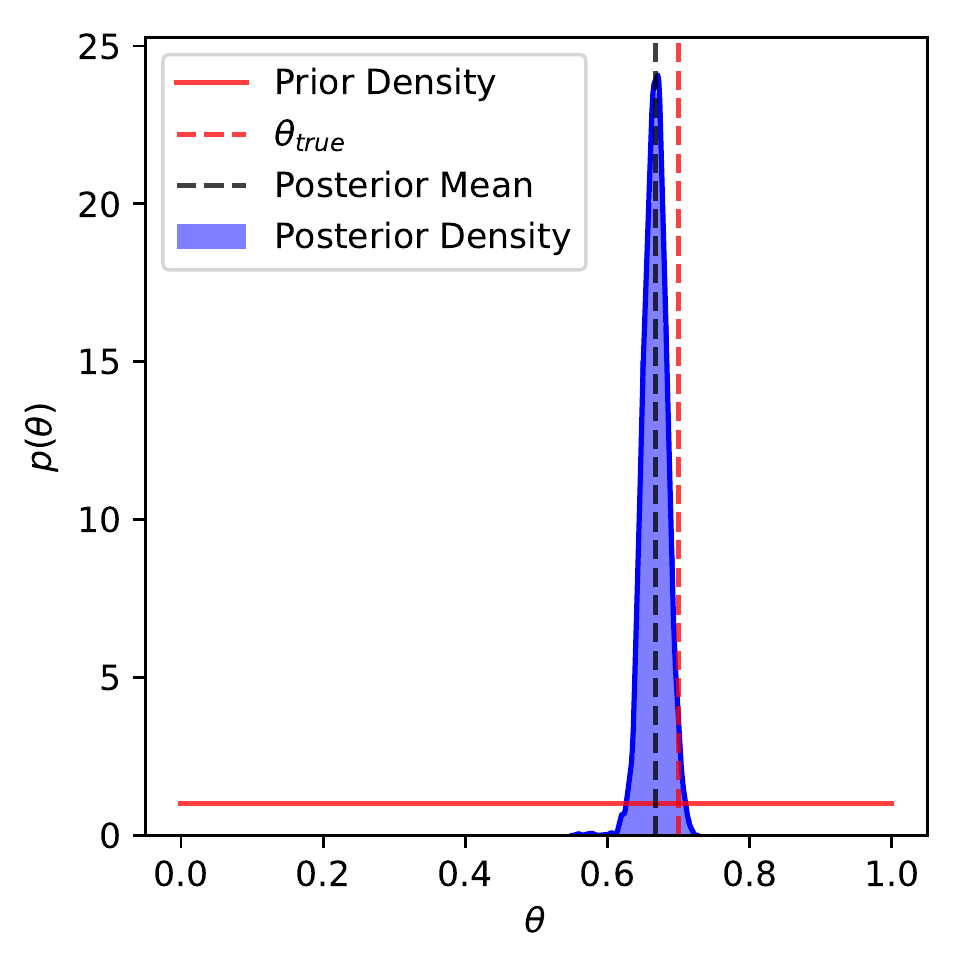}
	\caption{Posterior distribution of parameter $a_1$.} \label{AR_1_Bayesian}
\end{subfigure}

\caption{A graphical illustration of the calibration results obtained for the AR$(1)$ model.} 

\end{figure}

Referring to Figure \ref{AR_1_Curves}, where we plot the objective function curves for each of the considered SMD methods and the corresponding minima found by each optimisation algorithm, we find that, as expected, each method produces estimates close to the true parameter value with little difficulty. Similarly, when referring to Figure \ref{AR_1_Bayesian}, where we present the posterior distribution of $a_1$ obtained using Bayesian estimation, we find that the posterior mean (our selected point estimate) is indeed close to the true parameter value and is estimated with a relatively small degree of uncertainty, as is evident from the low variance implied by the obtained density function.

More detailed results, including the loss function values associated with each point estimate, are presented in Table \ref{AR_1_Results_Table}, with the loss function suggesting that Bayesian estimation delivers the best performance for the simple AR$(1)$ model, followed by the MIC, MSM, and finally the GSL-div.

\subsubsection{ARMA$(2, 2)$-ARCH$(2)$ Model}

The next model to be calibrated, an ARMA$(2, 2)$-ARCH$(2)$ model, has a larger parameter space and is capable of producing more complex dynamics than the AR$(1)$ model. It therefore warrants a more comprehensive investigation, leading us to consider two free parameter sets, $[a_0, a_1]$ and $[b_1, b_2, c_0, c_1, c_2]$, where all parameters are assumed to lie in the interval $[0, 1]$, with the exception of $a_1$, which we assume lies in the interval $[0, 0.8]$\footnote{Values of $a_1 > 0.8$ can lead to the model becoming non-stationary.}.

\begin{figure}[H]

\centering

\begin{subfigure}{0.45\textwidth}
	\centering
	\includegraphics[width=1\linewidth]{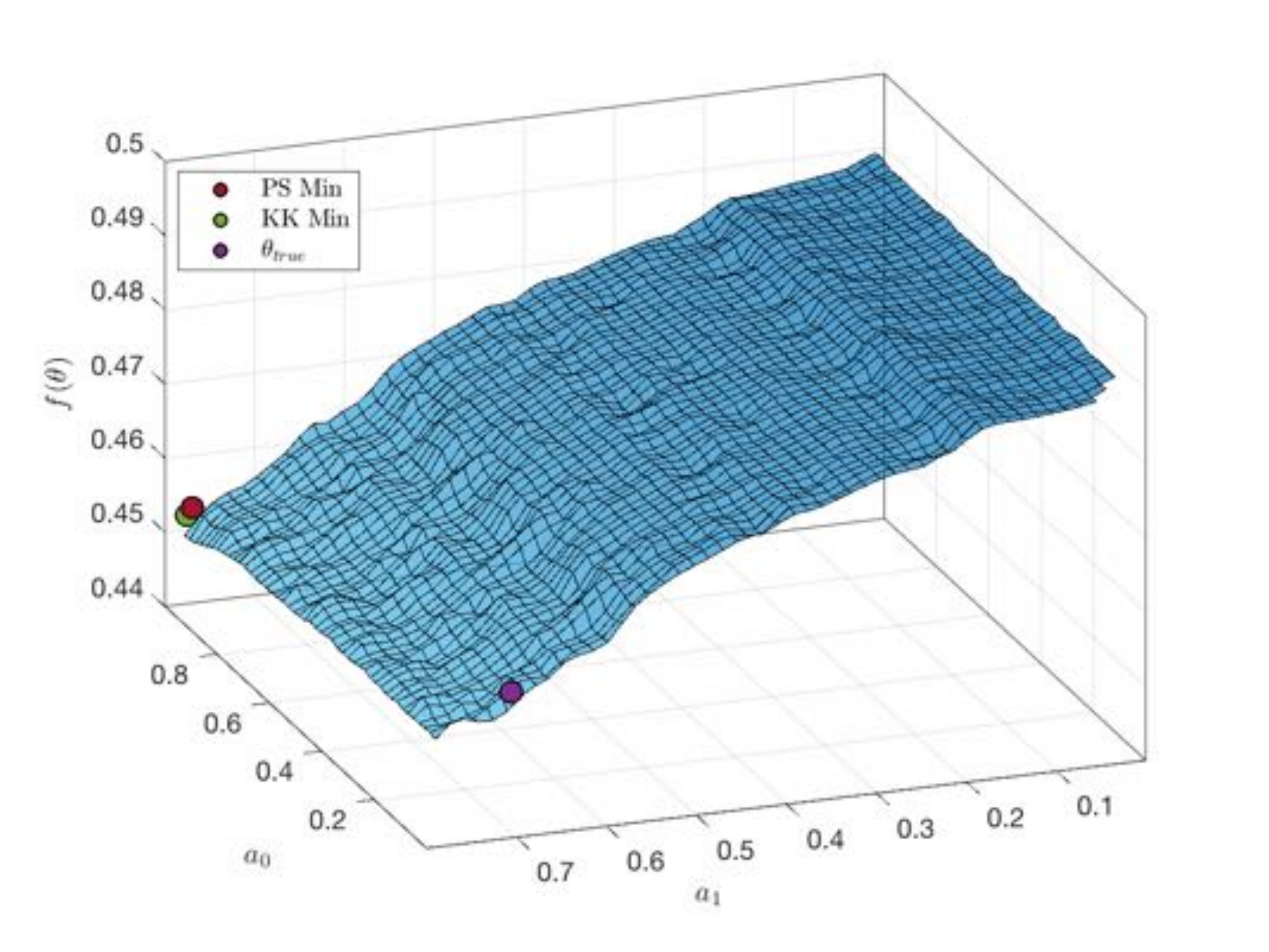}
	\caption{GSL-div}
\end{subfigure}%
\begin{subfigure}{0.45\textwidth}
	\centering
	\includegraphics[width=1\linewidth]{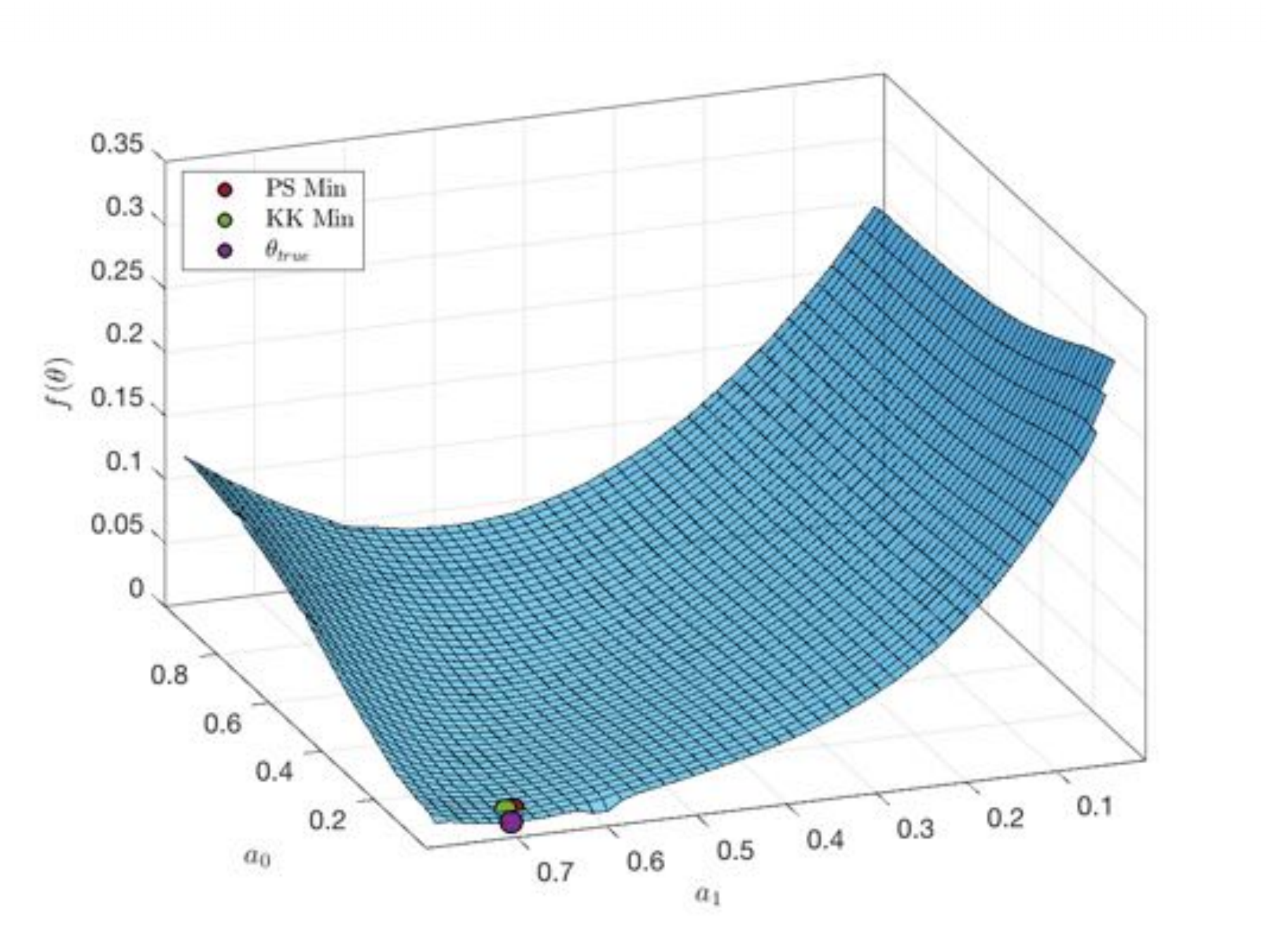}
	\caption{MSM}
\end{subfigure}
\begin{subfigure}{0.45\textwidth}
	\centering
	\includegraphics[width=1\linewidth]{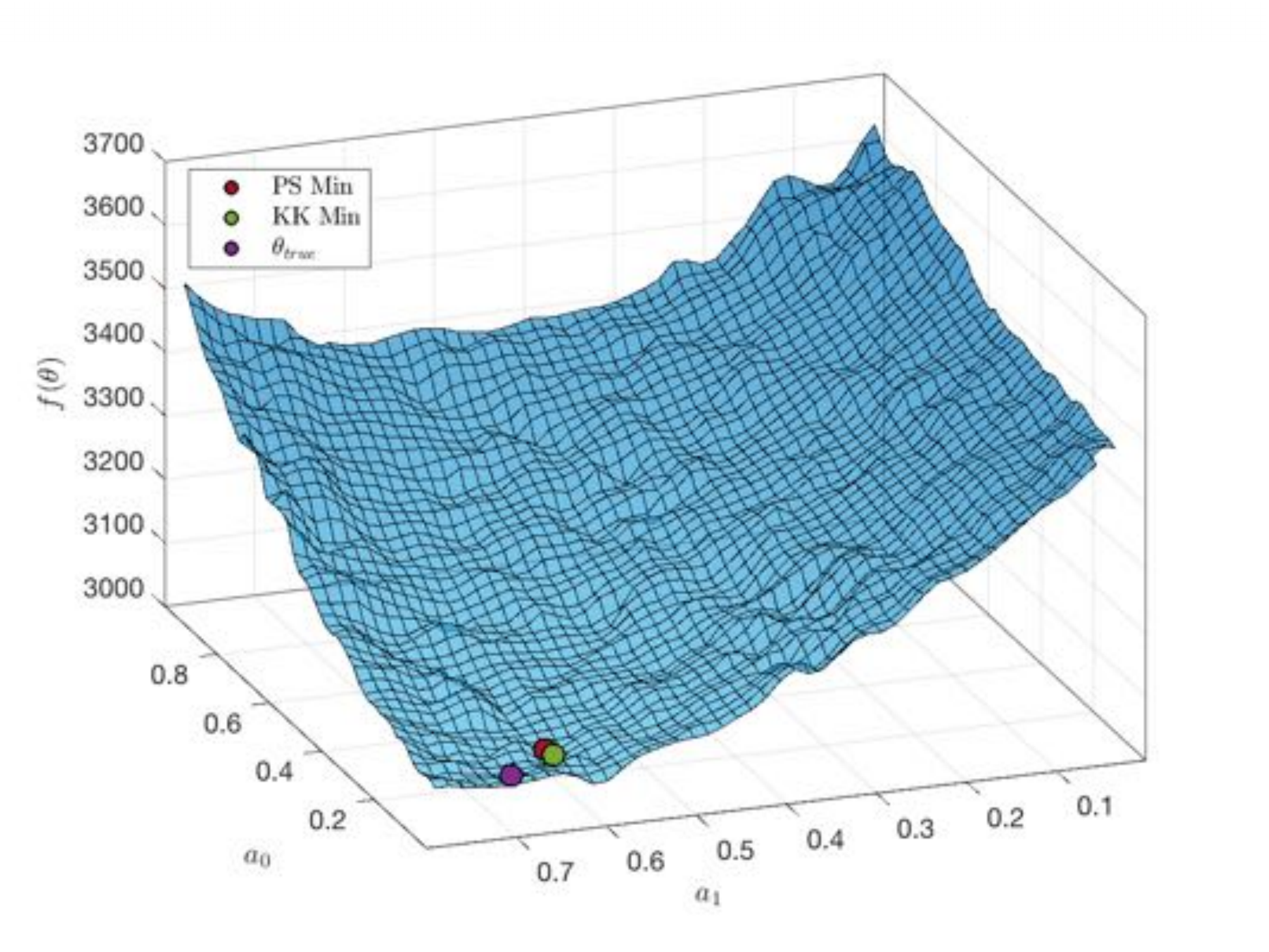}
	\caption{MIC}
\end{subfigure}

\caption{SMD objective function surfaces for free parameter set $1$ of the ARMA$(2, 2)$-ARCH$(2)$ model.} \label{ARMA_ARCH_Surfaces}

\end{figure}

As is evident in Figure \ref{ARMA_ARCH_Surfaces}, where we plot the objective function surfaces associated with the first free parameter set, differences in the relative performance of the GSL-div, MSM and MIC have become more pronounced when attempting to calibrate this more sophisticated model. While MSM and the MIC produce reasonable parameter estimates, the GSL-div appears to have performed relatively poorly, yielding an estimate for $a_0$ that is significantly different from its true value. A more thorough inspection of the objective function surface reveals that changes in the value of $a_0$ appear to have a limited effect on the GSL-div. The fact that $a_0$ is simply an additive constant suggests that the GSL-div is unable to differentiate between series of the form $\{x_t | t \geq 0\}$ and $\{x_t + C | t \geq 0, C \in \mathbb{R}\}$, an important limitation that is not yet discussed in the literature. 

\begin{figure}[H]

\centering

\includegraphics[width=0.8\linewidth]{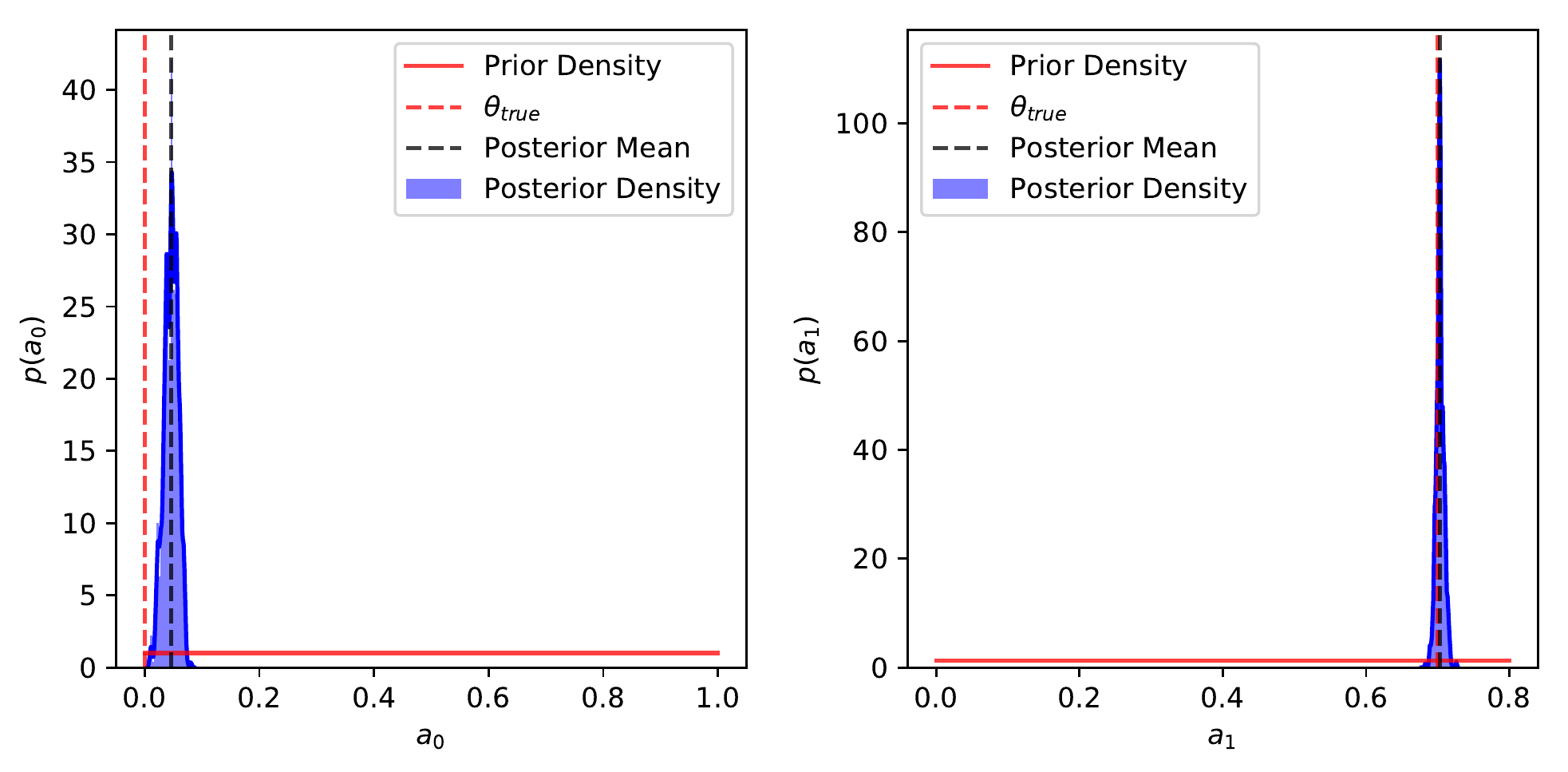}

\caption{Marginal posterior distributions for free parameter set $1$ of the ARMA$(2, 2)$-ARCH$(2)$ model.} \label{ARMA_ARCH_1_Bayesian}

\end{figure}

In the case of Bayesian estimation, we see that the method of \citet{Grazzini_et_al_2017} once again performs well, with the means of the posterior distributions of $a_0$ and $a_1$, shown in Figure \ref{ARMA_ARCH_1_Bayesian}, producing good estimates of the true parameter values. The loss function values presented in Table \ref{ARMA_ARCH_1_Results_Table} also suggest that Bayesian estimation is the best performing method, as was the case for the AR$(1)$ model, followed by MSM, the MIC and finally the GSL-div.

\begin{table}[h]

\caption{Calibration Results for Free Parameter Set $1$ of the ARMA$(2, 2)$-ARCH$(2)$ Model} \label{ARMA_ARCH_1_Results_Table}

\centering

\begin{tabularx}{\textwidth}{XXXXX}
\hline
& $a_0$ & $a_1$ & $L(\hat{\bm{\theta}}, \bm{\theta}_{true})$\\
\hline
$\bm{\theta}_{true}$ & $0$ & $0.7$ & $0$ \\
GSL-div/PS & $1.0$ & $0.8$ & $1.0050$ \\
GSL-div/KK & $1.0$ & $0.8$ & $1.0050$ \\
MSM/PS & $0.0694$ & $0.6831$ & $0.0715$ \\
MSM/KK & $0.069$ & $0.6914$ & $0.0695$ \\
MIC/PS & $0.13$ & $0.6297$ & $0.1478$ \\
MIC/KK & $0.0814$ & $0.6349$ & $0.1042$ \\
BE & $0.0459$ & $0.7033$ & $0.0460$ \\
\hline
\end{tabularx}

\end{table}

\begin{figure}[H]

\centerline{\includegraphics[width=0.9\linewidth]{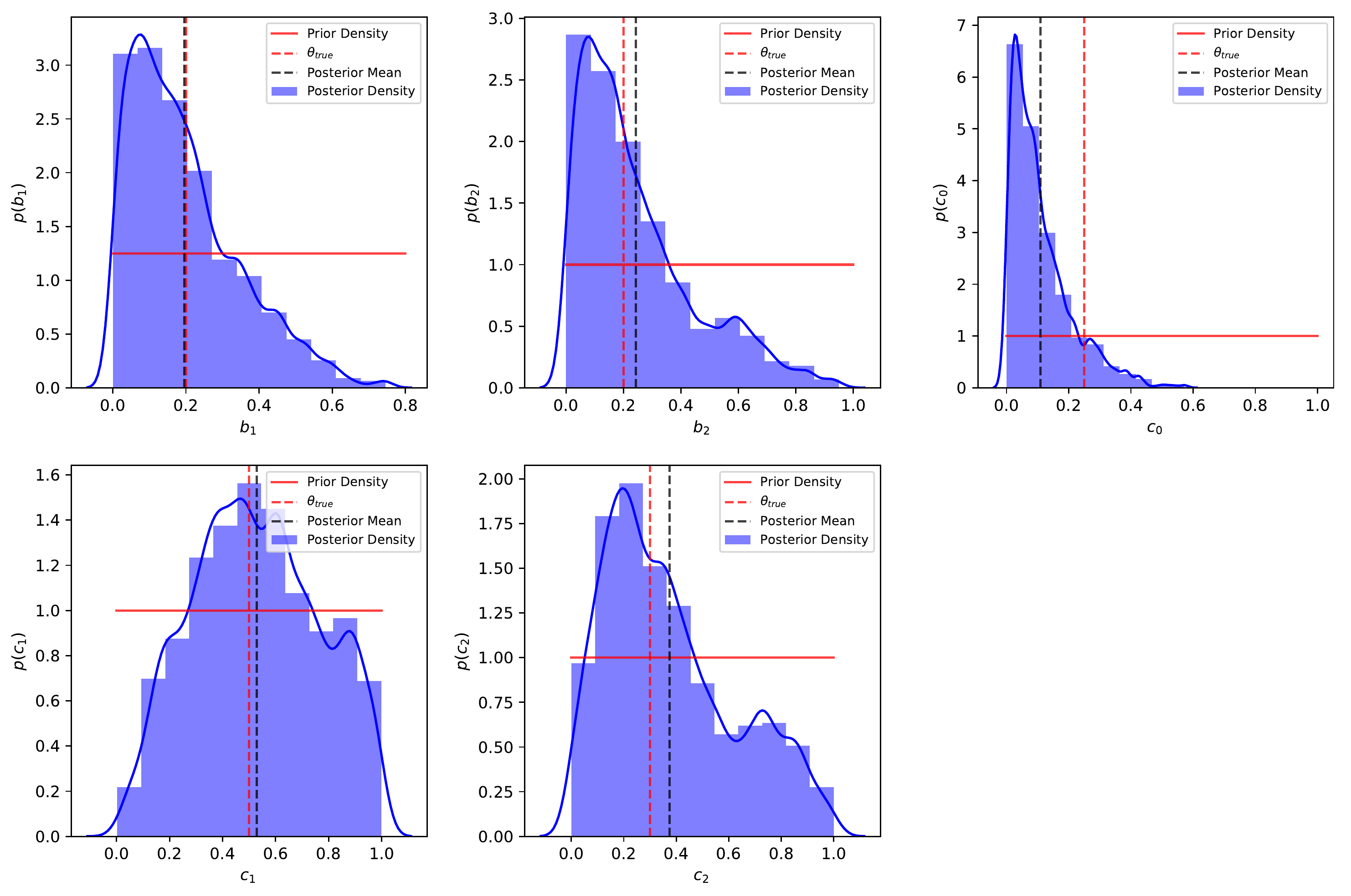}}

\caption{Marginal posterior distributions for free parameter set $2$ of the ARMA$(2, 2)$-ARCH$(2)$ model.} \label{ARMA_ARCH_2_Bayesian}

\end{figure}

At this point, it is natural to ask whether similar behaviours are observed for free parameter sets of greater cardinality. Referring to Figure \ref{ARMA_ARCH_2_Bayesian}, it is apparent that the performance of Bayesian estimation is indeed maintained in the more challenging context of $5$ free parameters, with the resultant estimates again being reasonably close to the true parameter values\footnote{Note that since our calibration experiments involve the comparison of an ensemble of Monte Carlo replications, all of which are generated using different random seeds, to a single series, itself generated using a unique random seed, the observed level of noise is to be expected.}. Nevertheless, the uncertainty of estimation is greater than in the previously presented cases, as suggested by the increased variance of the posterior distributions obtained for the second free parameter set. This is to be expected, however, since the number of parameter combinations that produce a reasonable fit to a given dataset is likely to increase with the number of free parameters, which could most likely be addressed by the consideration of additional data or a larger number of Monte Carlo replications\footnote{Note that this ability to quantify the uncertainty of estimation, albeit rather crudely, offers a distinct advantage over SMD methods, which simply produce point estimates and lack an inherent method of deducing the degree of uncertainty without repeated estimation attempts on multiple (bootstrapped) datasets.}.

In contrast to Bayesian estimation, SMD methods do not seem to maintain a similar level of performance as the number of dimensions is increased, as indicated by the significant differences between the obtained estimates and true parameter values presented in Table \ref{ARMA_ARCH_2_Results_Table}.  One should also be aware that the optimisation algorithms no longer agree on the obtained estimates, which can be seen as indicative of convergence to local minima. 

\begin{table}[h]

\caption{Calibration Results for Free Parameter Set $2$ of the ARMA$(2, 2)$-ARCH$(2)$ Model} \label{ARMA_ARCH_2_Results_Table}

\centering

\begin{tabularx}{\textwidth}{lXXXXXXX}
\hline
& $b_1$ & $b_2$ & $c_0$ & $c_1$ & $c_2$ & $L(\hat{\bm{\theta}}, \bm{\theta}_{true})$\\
\hline
$\bm{\theta}_{true}$ & $0.2$ & $0.2$ & $0.25$ & $0.5$ & $0.3$ & $0$ \\
GSL-div/PS & $0.1764$ & $0.0546$ & $0.7073$ & $0$ & $0$ & $0.7555$ \\
GSL-div/KK & $0.1707$ & $0$ & $1$ & $0$ & $0$ & $0.9713$ \\
MSM/PS & $0.0592$ & $0.2964$ & $0.0$ & $0.0506$ & $1$ & $0.8852$ \\
MSM/KK & $0.0962$ & $0.2323$ & $0.0477$ & $0.23$ & $0.8471$ & $0.6519$ \\
MIC/PS & $0.5389$ & $0$ & $0$ & $0$ & $0$ & $0.7466$ \\
MIC/KK & $0$ & $1$ & $0$ & $0$ & $0$ & $1.0404$ \\
BE & $0.1959$ & $0.2432$ & $0.109$ & $0.5283$ & $0.3743$ & $0.1676$ \\
\hline
\end{tabularx}

\end{table}

A possible explanation for the observed phenomena is that the objective functions are characterised by the presence of large, flat regions of local minima (possibly containing the true parameter values) as opposed to a single global minimum. This leads to parameter identification difficulties and the obtained minima being dependent on the initial conditions of the optimisation algorithms. This is also consistent with the results obtained using Bayesian estimation, which suggest that there are potentially many feasible parameter sets capable of producing a reasonable fit to the data. It would appear, however, that the mean of the joint posterior distribution produces a far better estimate of the true parameter values than the minimisation of objective functions at a single point.

\subsubsection{Random Walks with Structural Breaks}

While the ARMA$(2, 2)$-ARCH$(2)$ model is indeed capable of producing significantly more nuanced dynamics than the AR$(1)$ model, it is still unable to replicate a range of behaviours observed in both real economic data and the outputs of large-scale ABMs. Among these phenomena, structural breaks, or sudden and dramatic deviations from prior temporal trends, are of particular interest. This leads us to investigate the extent to which a random walk model capable of producing simple structural breaks can be successfully calibrated using the considered methods.

\begin{figure}[H]

\centering

\begin{subfigure}{1\textwidth}
	\centering
	\includegraphics[width=1\linewidth]{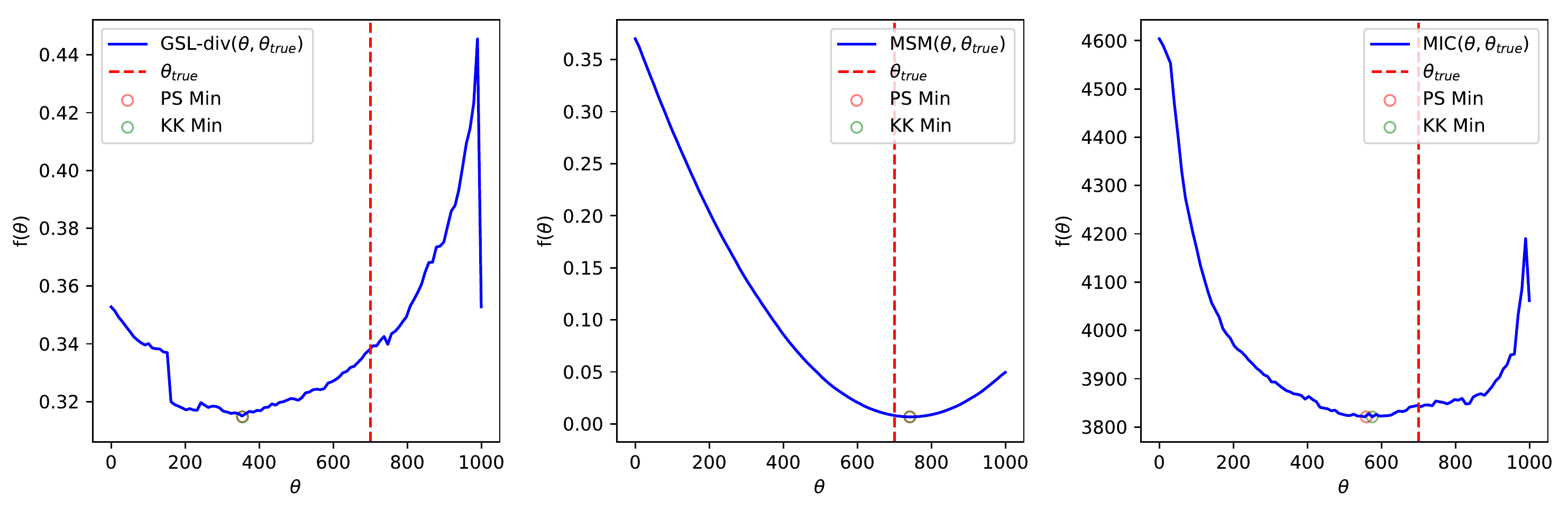}
	\caption{SMD objective function curves for parameter $\tau$.} \label{RW_1_Curves}
\end{subfigure}

\vspace{0.45cm}

\begin{subfigure}{1\textwidth}
	\centering
	\includegraphics[width=0.32\linewidth]{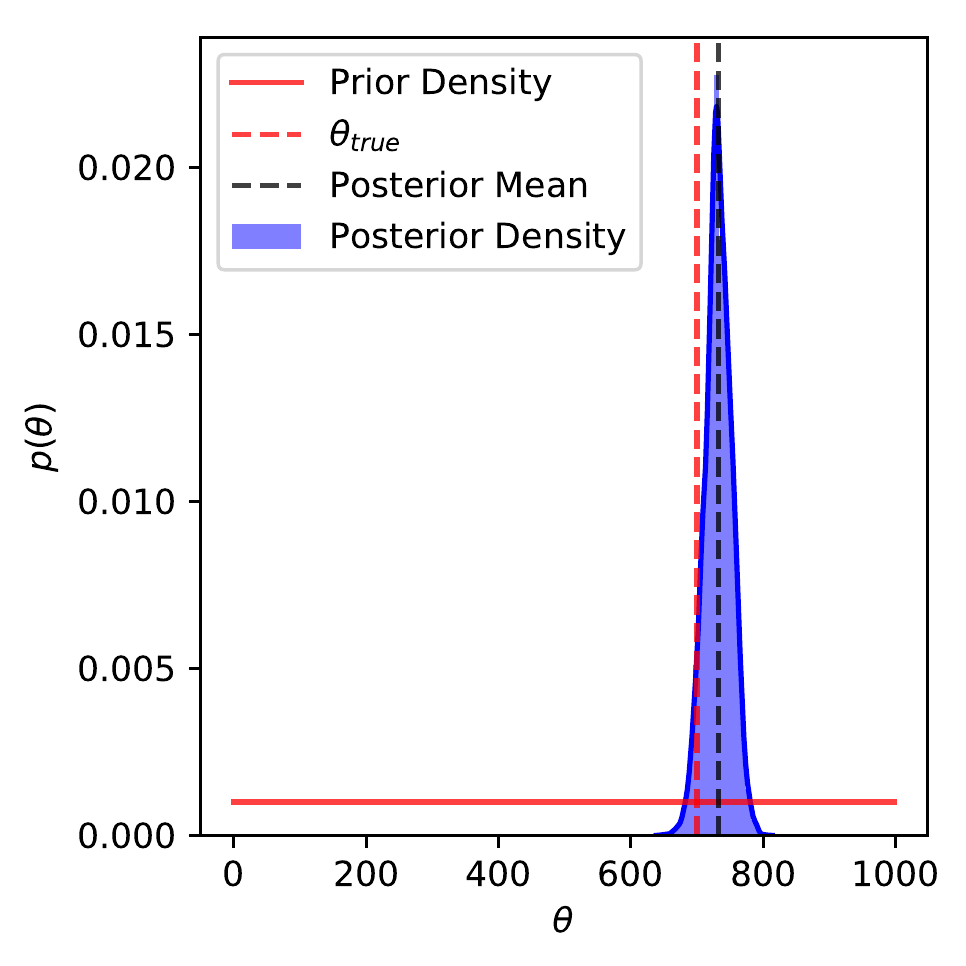}
	\caption{Posterior distribution of parameter $\tau$.} \label{RW_1_Bayesian}
\end{subfigure}

\caption{A graphical illustration of the calibration results obtained for parameter $\tau$ of the random walk model.} 

\end{figure}

In this simplified context, a successful method would not only be able to calibrate parameters which determine the pre- and post-break dynamics, but should also be able to identify the point at which the structural break occurs, even if not precisely. We therefore consider two free parameter sets, $[\tau] \in [0, 1000]$ and $[\sigma_1, \sigma_2], \sigma_{i} \in [0, 1]$, which correspond to each of the aforementioned aspects.

Beginning with Figure \ref{RW_1_Curves}, which presents the SMD objective function curves resulting from attempts to determine the location of the structural break, we see that both MSM and the MIC are capable of inferring the correct location to some extent, although the estimate associated with MSM is far closer to the true value. The GSL-div, on the other hand, delivers far less compelling results, with its associated objective function curve suggesting that values roughly in the interval $[180, 700]$, over half of the total parameter range, are preferable to the true value. 

Moving away from SMD methods, Figure \ref{RW_1_Bayesian} illustrates the ability of Bayesian estimation to estimate $\tau$ to a reasonable extent and with a relatively low degree of uncertainty. Furthermore, the loss function values presented in Table \ref{RW_1_Results_Table} suggest that Bayesian estimation again delivers the best performance among the considered methods.

\begin{table}[h]

\caption{Calibration Results for Parameter $\tau$ of the Random Walk Model} \label{RW_1_Results_Table}

\centering

\begin{tabularx}{\textwidth}{XXX}
\hline
& $\tau$ & $L(\hat{\bm{\theta}}, \bm{\theta}_{true})$\\
\hline
$\bm{\theta}_{true}$ & $700$ & $0$ \\
GSL-div/PS & $354$ & $346$ \\
GSL-div/KK & $355$ & $355$ \\
MSM/PS & $741$ & $41$ \\
MSM/KK & $742$ & $42$ \\
MIC/PS & $559$ & $141$ \\
MIC/KK & $575$ & $125$ \\
BE & $732$ & $32$ \\
\hline
\end{tabularx}

\end{table}

At this point, one can identify the emergence of a persistent trend that appears throughout the preceding experiments. In particular, notice that Bayesian estimation always delivers the best calibration performance (as measured by the loss function) and has, in all of the cases considered thus far, been relatively successful in recovering the parameters used to generate the artificial datasets. In contrast to this, SMD methods are less consistent in their performance, with some methods performing well in certain contexts and poorly in others. In most cases, however, it would appear that MSM and the MIC perform better than the GSL-div.

\begin{figure}[H]

\centering

\includegraphics[width=0.8\linewidth]{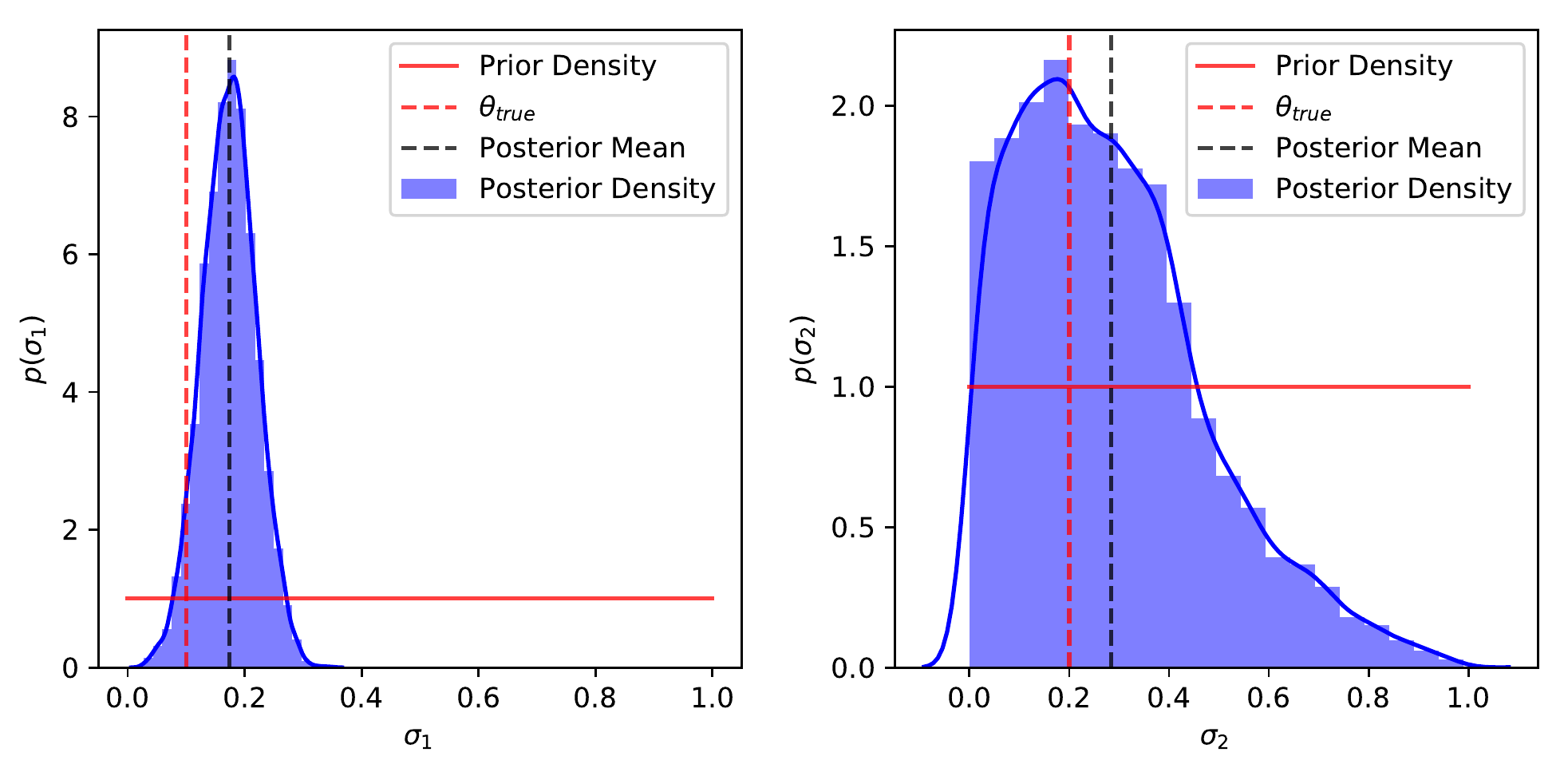}

\caption{Marginal posterior distributions for free parameter set $2$ of the random walk model.} \label{RW_2_Bayesian}

\end{figure}

\begin{figure}[H]

\centering

\begin{subfigure}{0.45\textwidth}
	\centering
	\includegraphics[width=1\linewidth]{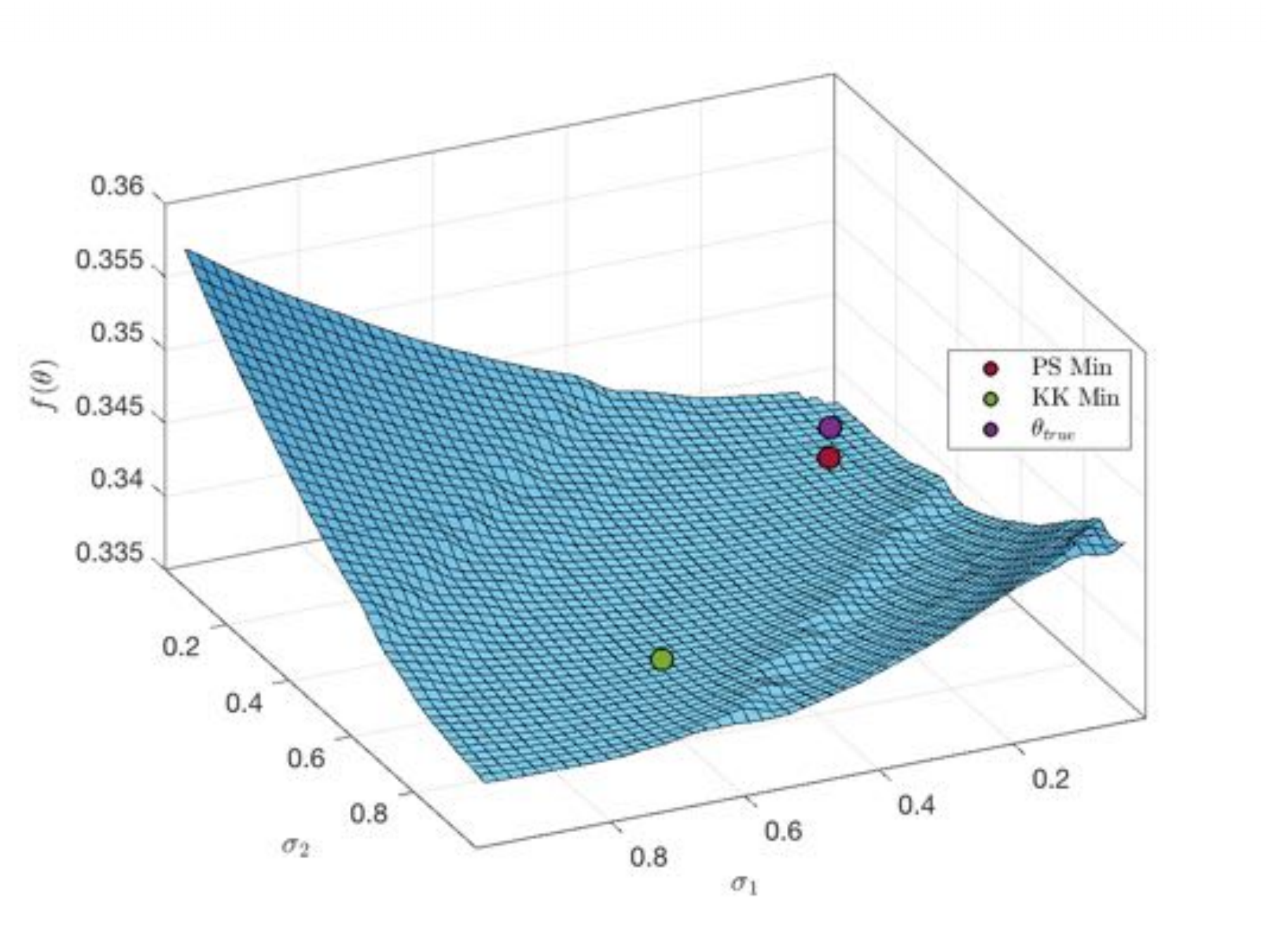}
	\caption{GSL-div}
\end{subfigure}%
\begin{subfigure}{0.45\textwidth}
	\centering
	\includegraphics[width=1\linewidth]{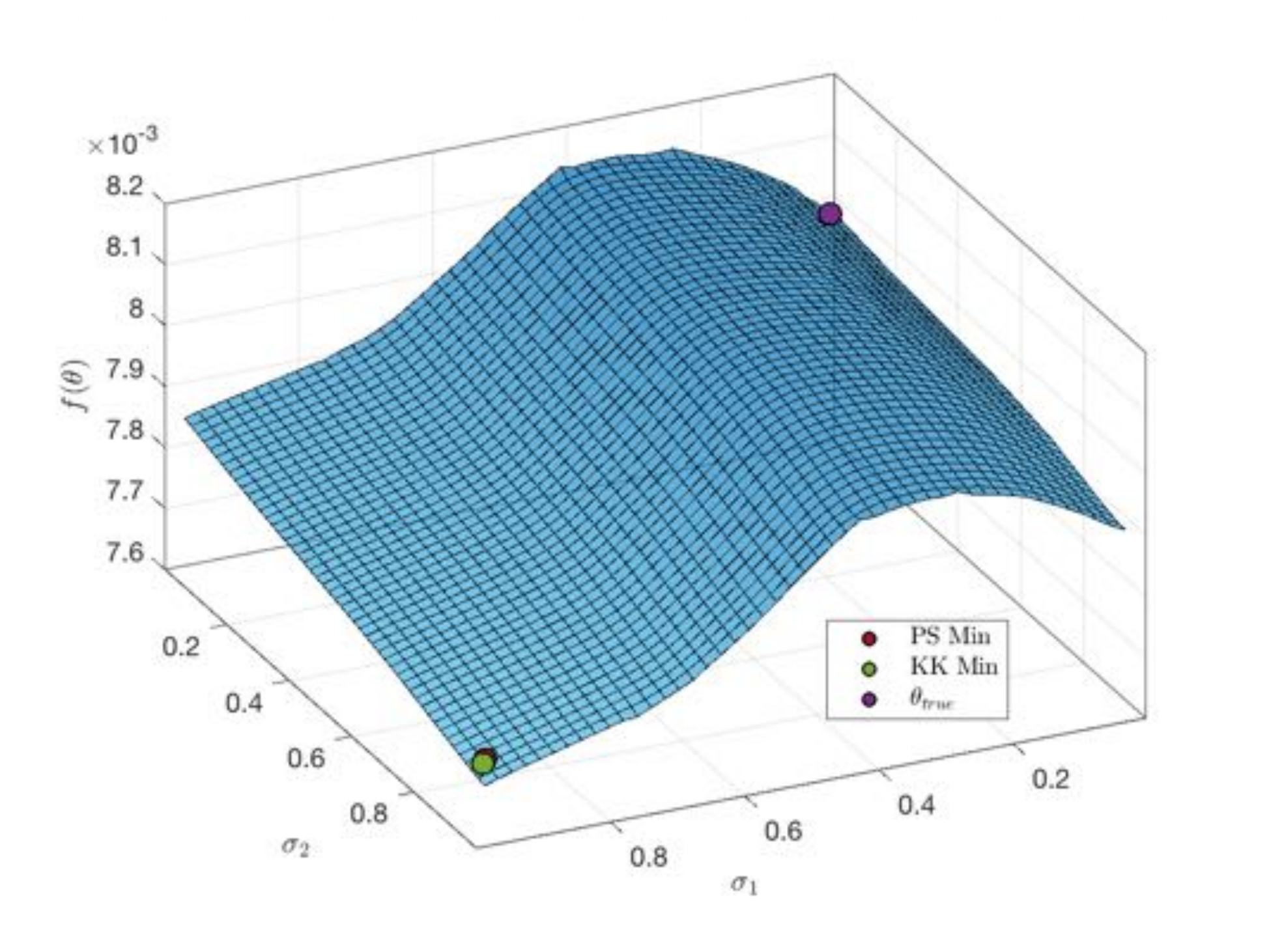}
	\caption{MSM}
\end{subfigure}
\begin{subfigure}{0.45\textwidth}
	\centering
	\includegraphics[width=1\linewidth]{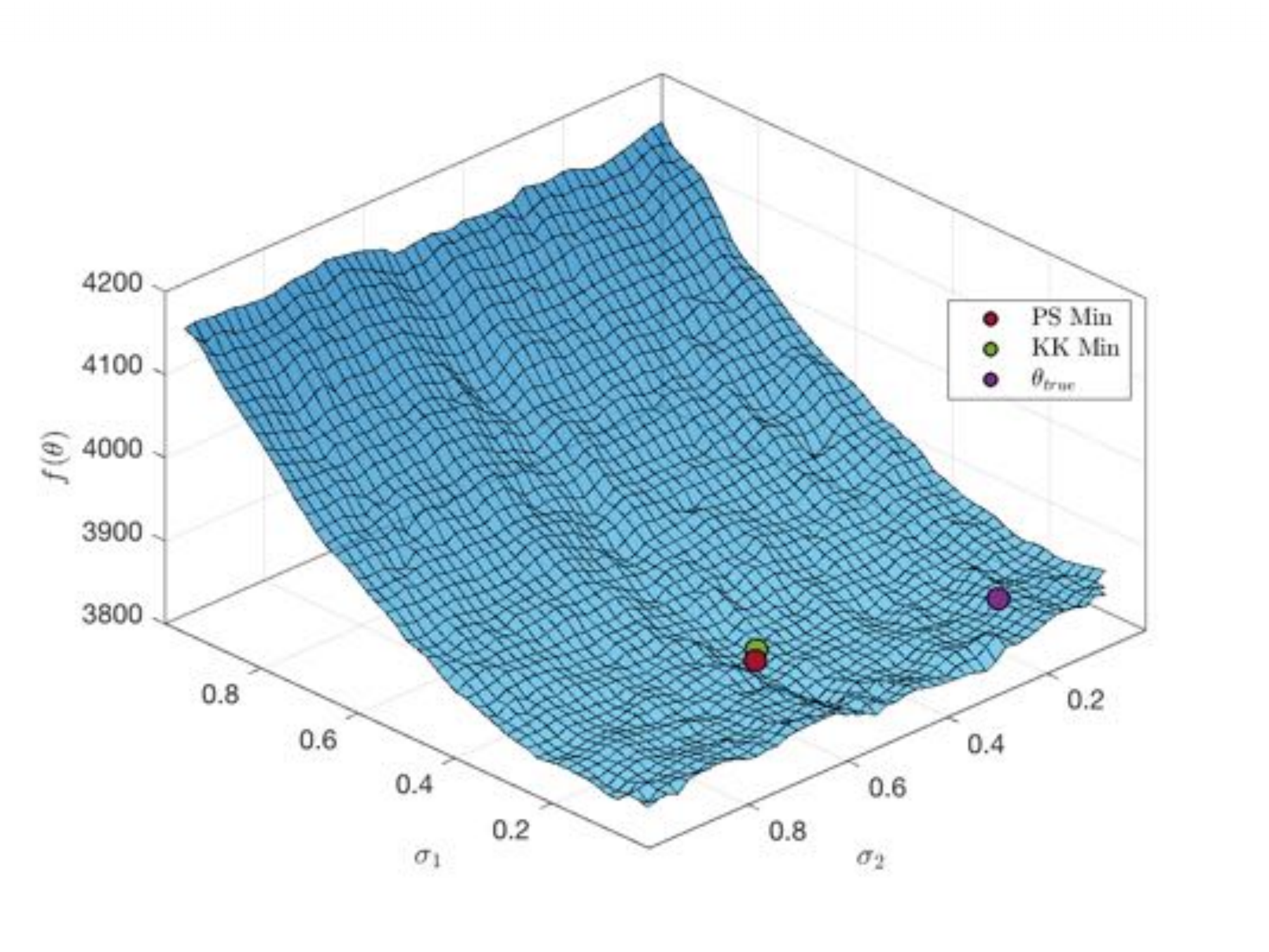}
	\caption{MIC}
\end{subfigure}

\caption{SMD objective function surfaces for free parameter set $2$ of the random walk model.} \label{RW_2_Surfaces}

\end{figure}

Unsurprisingly, Figures \ref{RW_2_Bayesian} and \ref{RW_2_Surfaces} demonstrate that this trend largely persists in the case of the second free parameter set. Note, however, that the loss function values presented in Table \ref{RW_2_Results_Table} suggest that the GSL-div minimum obtained using particle swarm optimisation is a better estimate of the true parameter values than the posterior mean obtained using Bayesian estimation, while, paradoxically, the GSL-div minimum obtained using the method of \citet{Knysh_Korkolis_2016} is a significantly worse estimate. A closer examination of Figure \ref{RW_2_Surfaces} reveals that the GSL-div objective function surface includes a large, flat region containing the true parameter values. Therefore, the proximity of the estimate obtained using particle swarm optimisation to the true parameter values seems to simply be a fortunate coincidence of the initialisation of the optimisation algorithm as opposed to a robust result.

Recall that a similar lack of agreement between the minima obtained by the considered optimisation algorithms was observed for the second free parameter set of the ARMA$(2, 2)$-ARCH$(2)$ model, which we suggested may also stem from large, flat regions being present in the objective functions. The visual presence of the flat region in Figure \ref{RW_2_Surfaces} provides some evidence that this may indeed be the case.

\begin{table}[h]

\caption{Calibration Results for Free Parameter Set $2$ of the Random Walk Model} \label{RW_2_Results_Table}

\centering

\begin{tabularx}{\textwidth}{XXXX}
\hline
& $\sigma_1$ & $\sigma_1$ & $L(\hat{\bm{\theta}}, \bm{\theta}_{true})$\\
\hline
$\bm{\theta}_{true}$ & $0.1$ & $0.2$ & $0$ \\
GSL-div/PS & $0.1426$ & $0.2875$ & $0.0973$ \\
GSL-div/KK & $0.6238$ & $0.7879$ & $0.7874$ \\
MSM/PS & $1$ & $1$ & $1.2042$ \\
MSM/KK & $1$ & $1$ & $1.2042$ \\
MIC/PS & $0.2325$ & $0.5599$ & $0.3835$ \\
MIC/KK & $0.26$ & $0.5307$ & $0.3674$ \\
BE & $0.1737$ & $0.2835$ & $0.1114$ \\
\hline
\end{tabularx}

\end{table}

Finally, notice that although the optimisation algorithms are able to identify the global minima in the case of the MIC and MSM, the obtained minima, presented in Table \ref{RW_2_Results_Table}, are not consistent with the true parameter set. This suggests that, in the case of SMD methods, one has to be aware that even if a unique global minimum can be obtained in higher dimensions, it may not necessarily be a good estimate of the parameters of interest. A plausible explanation for this inconsistency is the possibility that the distance function may be unable to detect relevant aspects of the considered data. More concretely, MSM involves the estimation of various quantities for the full length of the time series and is therefore not well-suited to cases where different regimes characterise different regions of the data. In contrast to this, the MIC evaluates the binary log scores at the level of each observation, making it a more logical choice for problems of this nature. It is therefore no surprise that the loss function values associated with the MIC in Table \ref{RW_2_Results_Table} are nearly $4$ times smaller than those associated with MSM.

\subsubsection{\citet{Brock_Hommes_1998} Model}

\begin{figure}[H]

\centerline{\includegraphics[width=0.8\linewidth]{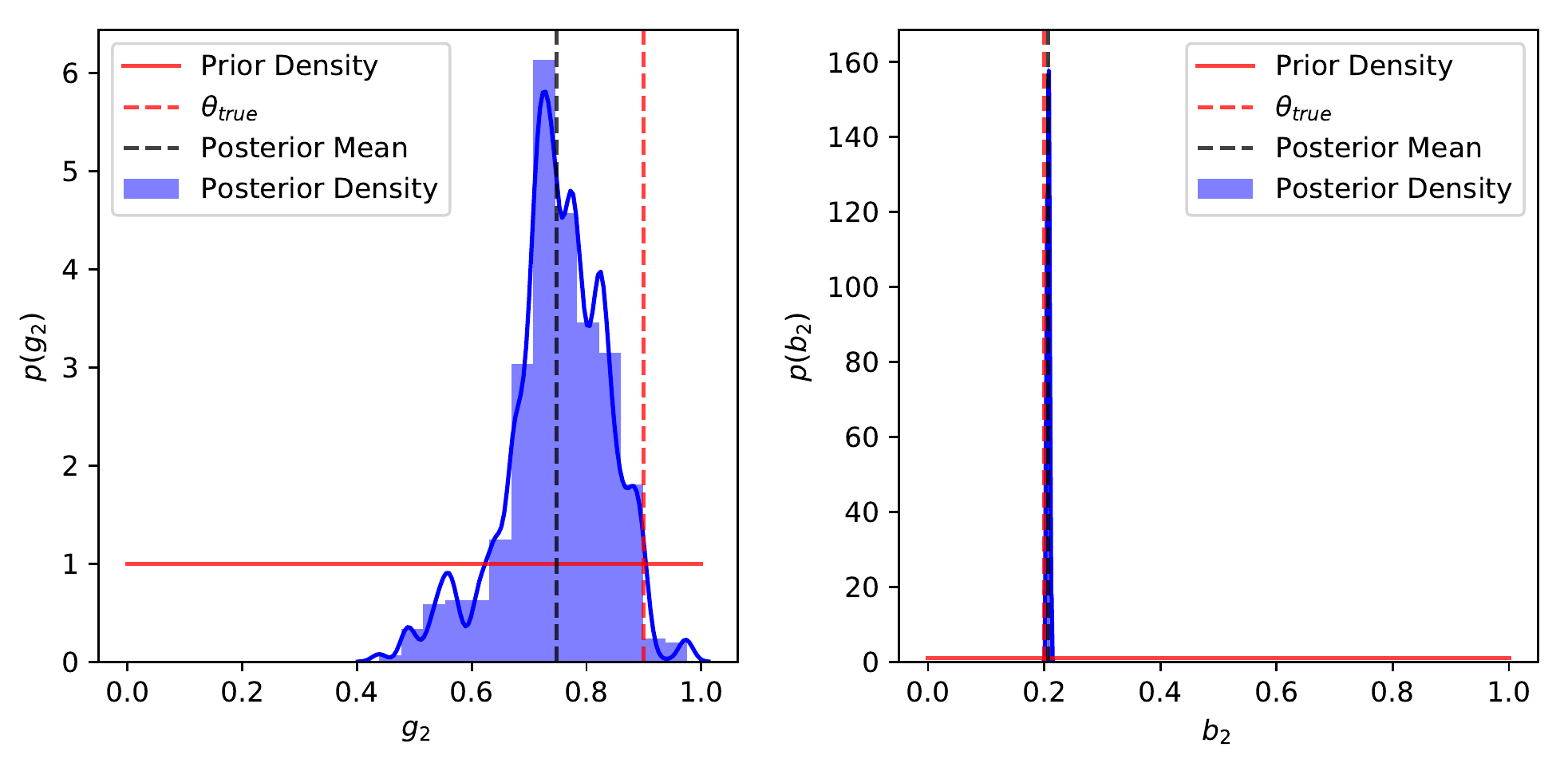}}

\caption{Marginal posterior distributions for free parameter set $1$ of the \citet{Brock_Hommes_1998} model.} \label{BH_1_Bayesian}

\end{figure}

\begin{figure}[H]

\centerline{\includegraphics[width=0.75\linewidth]{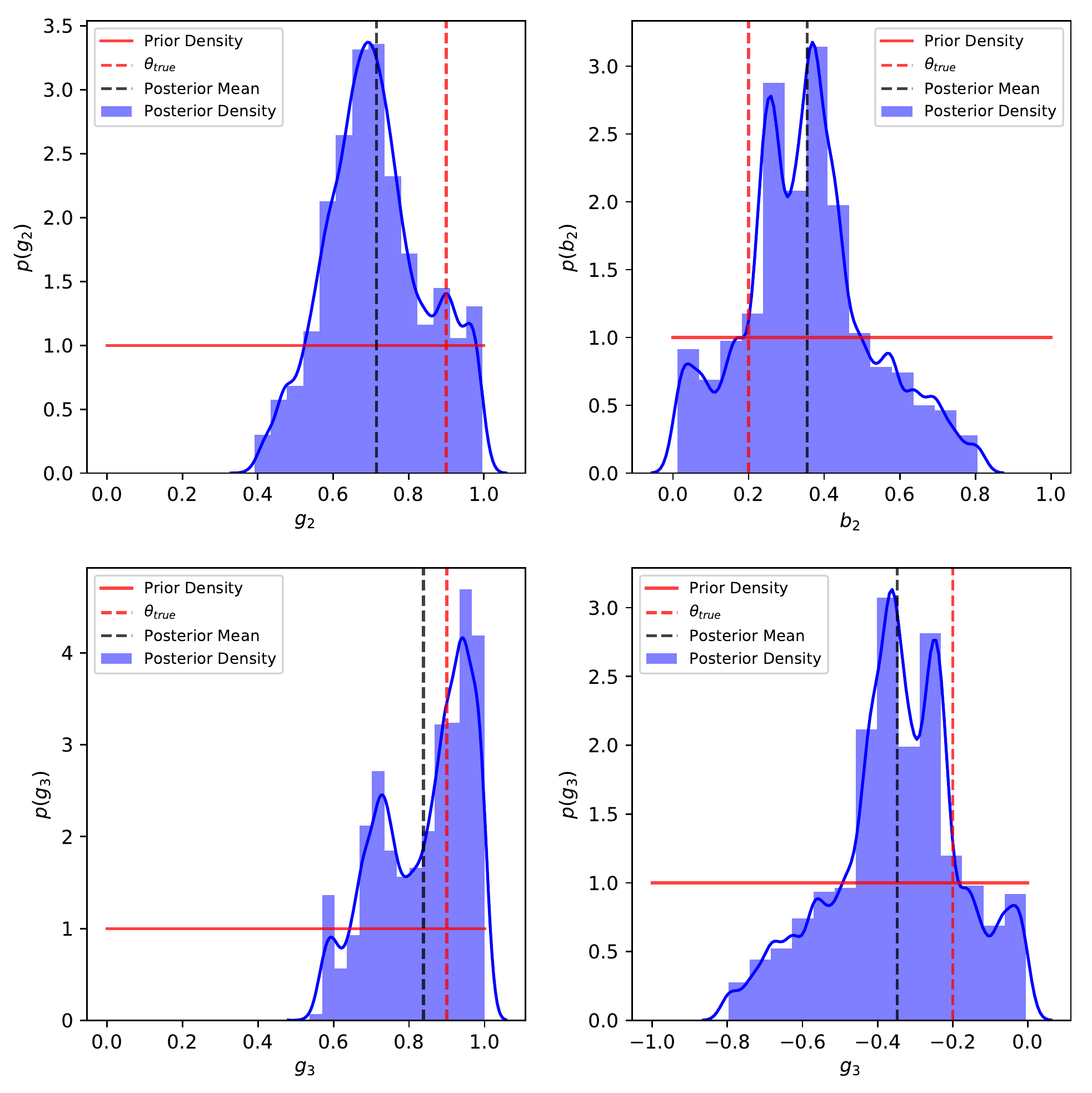}}

\caption{Marginal posterior distributions for free parameter set $2$ of the \citet{Brock_Hommes_1998} model.} \label{BH_2_Bayesian}

\end{figure}

\begin{figure}[H]

\centering

\begin{subfigure}{0.45\textwidth}
	\centering
	\includegraphics[width=1\linewidth]{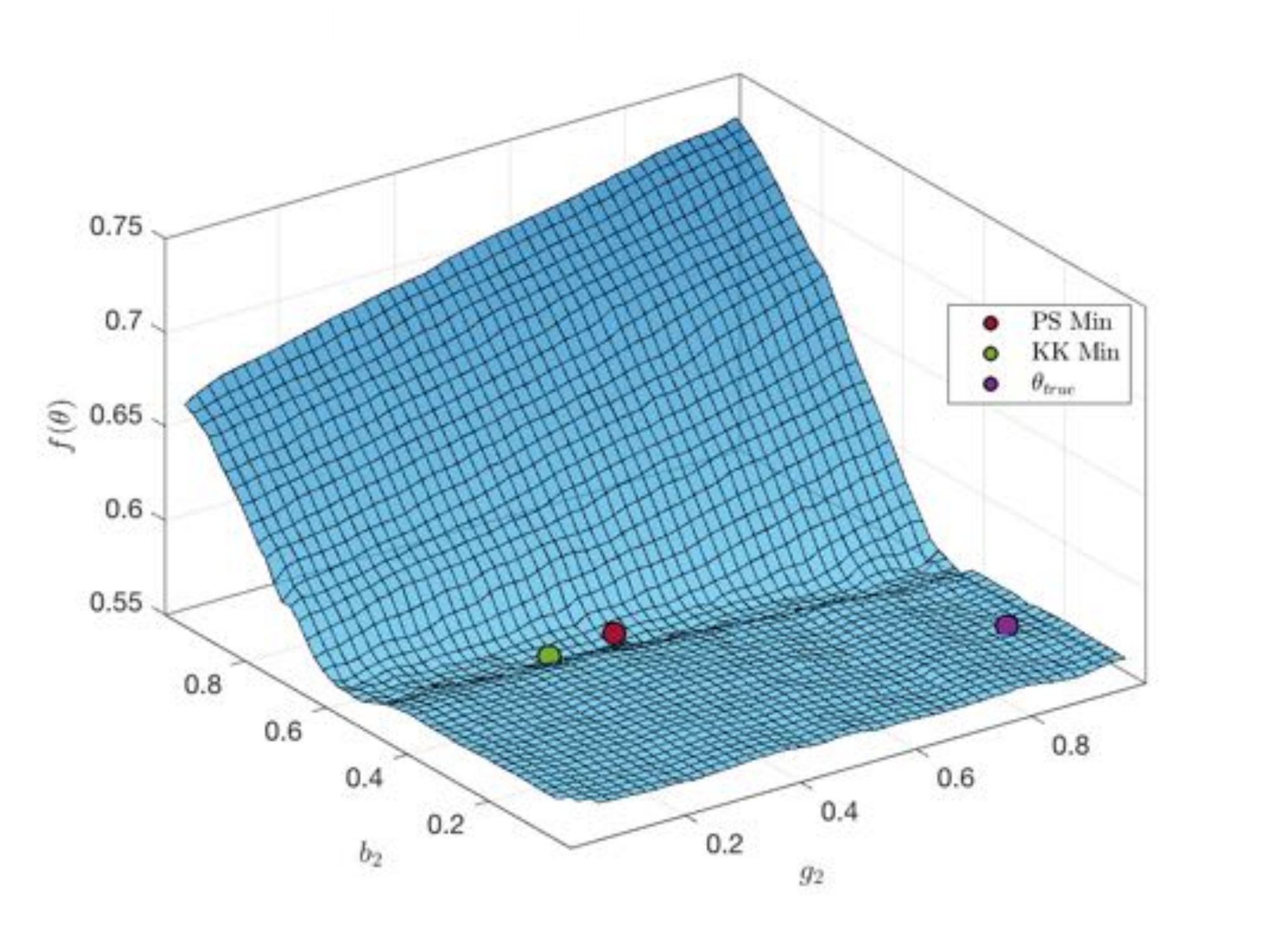}
	\caption{GSL-div}
\end{subfigure}%
\begin{subfigure}{0.45\textwidth}
	\centering
	\includegraphics[width=1\linewidth]{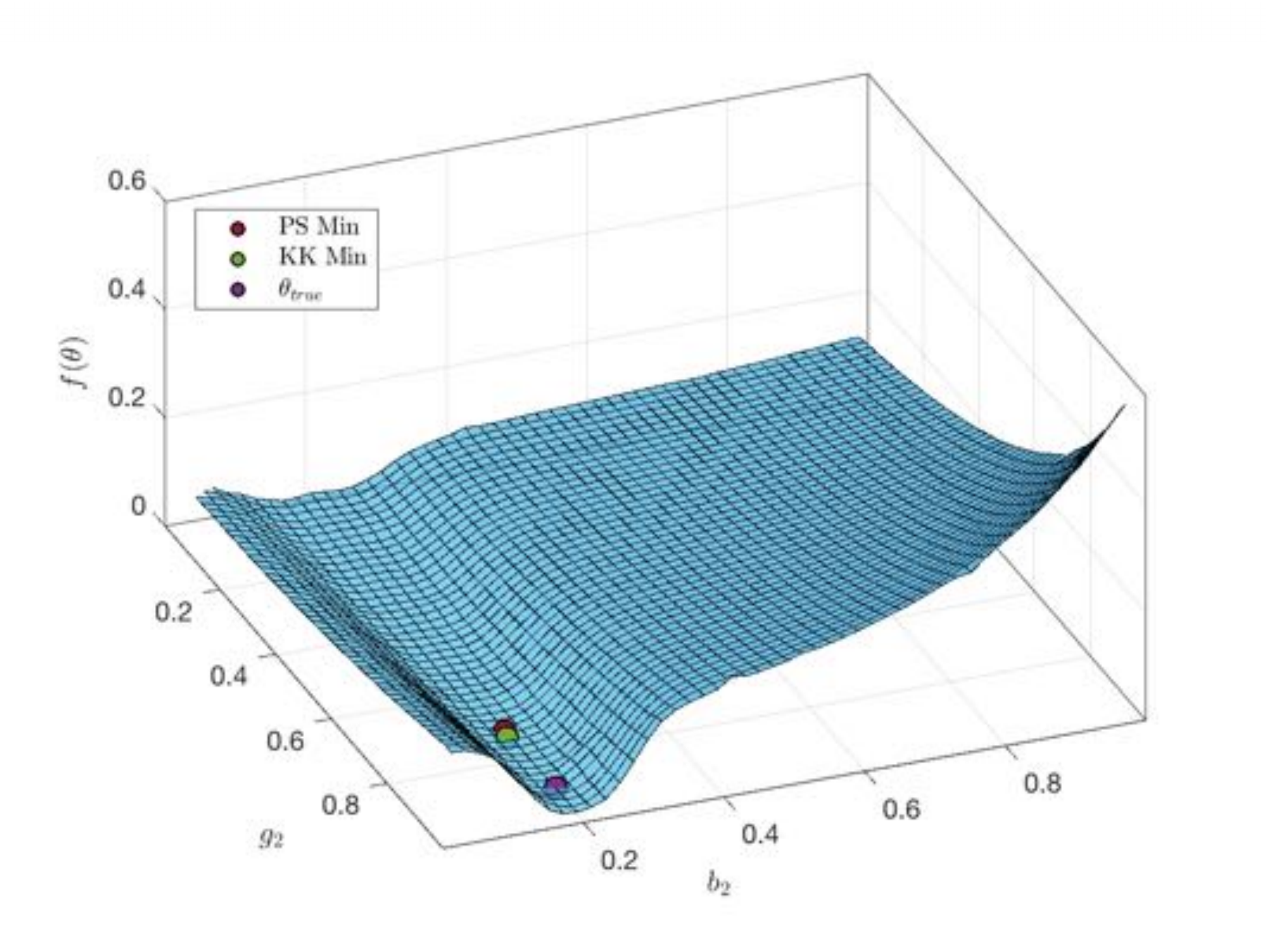}
	\caption{MSM}
\end{subfigure}
\begin{subfigure}{0.45\textwidth}
	\centering
	\includegraphics[width=1\linewidth]{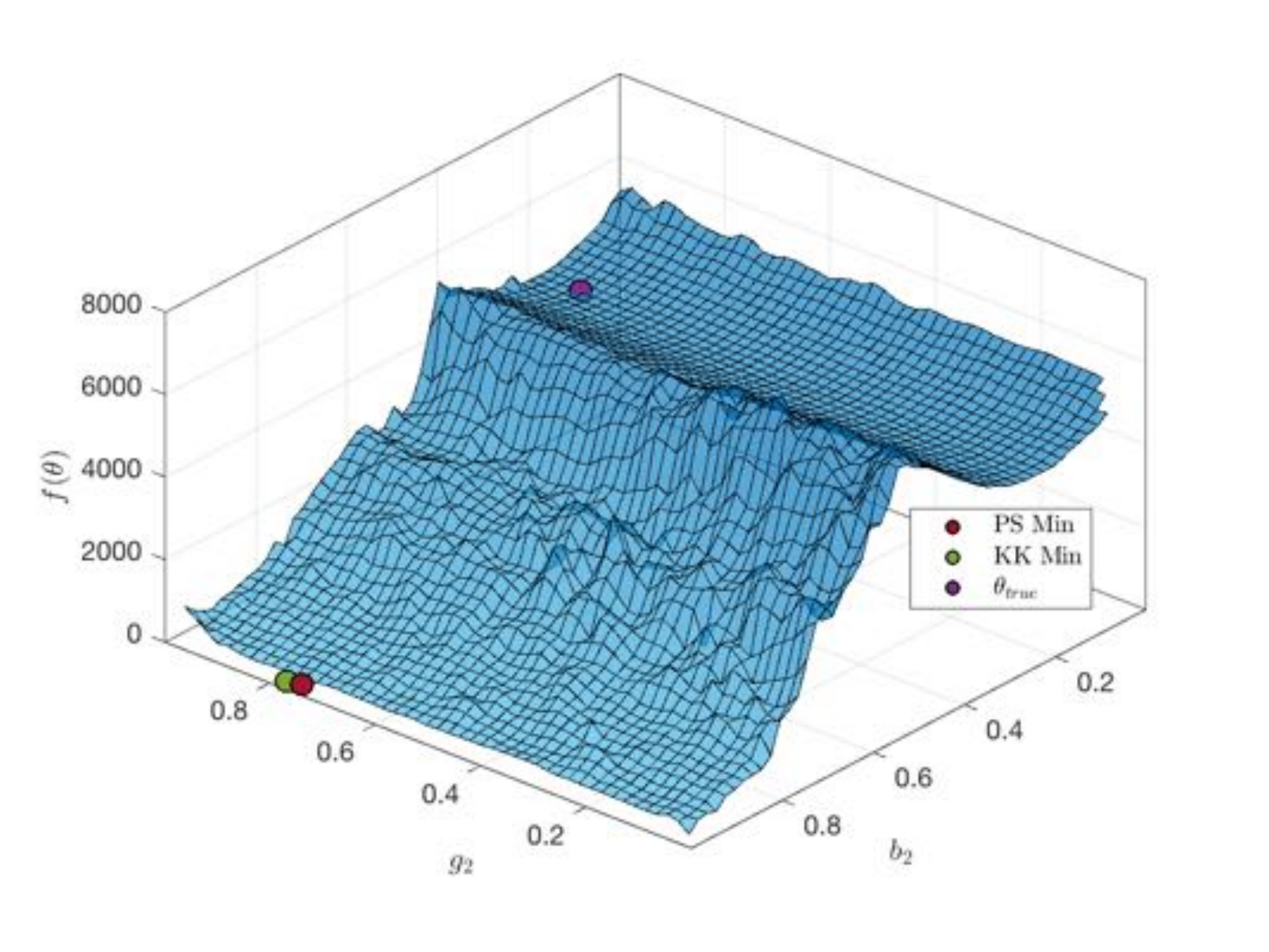}
	\caption{MIC}
\end{subfigure}

\caption{SMD objective function surfaces for free parameter set $1$ of the \citet{Brock_Hommes_1998} model.} \label{BH_1_Surfaces}

\end{figure}

The final model to be investigated is the \citet{Brock_Hommes_1998} model, considered a classic example of the use of heterogenous agents to model asset price time series. As previously stated, the model is ubiquitous in the calibration literature and is often used to test new approaches. Despite this, experiments of the type we consider here, or attempts to recover known parameter values, are still relatively rare. Given the size of the model's parameter space, there are potentially many free parameter sets of interest that could be considered, though we have ultimately settled on two, $[g_2, b_2]$ and $[g_2, b_2, g_3, b_3]$, with all parameters assumed to lie in the interval $[0, 1]$, with the exception of $b_3$, which we assume lies in the interval $[-1, 0]$.

As expected, the results obtained using Bayesian estimation, shown in Figures \ref{BH_1_Bayesian} and \ref{BH_2_Bayesian}, are similar to those observed for previous models, with the associated estimates being comparable to the true parameter values and the uncertainty of estimation increasing with the cardinality of the free parameter set. Similarly, the results obtained for free parameter set $1$ using SMD methods are also consistent with the previously identified trend that certain methods work well in some cases, but not in others. More concretely, Figure \ref{BH_1_Surfaces} indicates that MSM produces a reasonable estimate of the true parameter values, while this is not case for the MIC and GSL-div.

\begin{table}[H]

\caption{Calibration Results for Free Parameter Set $1$ of the \citet{Brock_Hommes_1998} Model} \label{BH_1_Results_Table}

\centering

\begin{tabularx}{\textwidth}{XXXX}
\hline
& $g_2$ & $b_2$ & $L(\hat{\bm{\theta}}, \bm{\theta}_{true})$\\
\hline
$\bm{\theta}_{true}$ & $0.9$ & $0.2$ & $0$ \\
GSL-div/PS & $0.4524$ & $0.5315$ & $0.5570$ \\
GSL-div/KK & $0.3316$ & $0.5204$ & $0.6524$ \\
MSM/PS & $0.7208$ & $0.1987$ & $0.1792$ \\
MSM/KK & $0.7412$ & $0.1944$ & $0.1589$ \\
MIC/PS & $0.7431$ & $0.9937$ & $0.8091$ \\
MIC/KK & $0.7725$ & $0.9935$ & $0.8037$ \\
BE & $0.7484$ & $0.2071$ & $0.1517$ \\
\hline
\end{tabularx}

\end{table}

Taking a final look at the constructed loss function, we see that the values presented in Table \ref{BH_1_Results_Table} indicate that Bayesian estimation is the best performing method for free parameter set $1$, as has been observed for all models and parameter sets considered up to this point. Rather surprisingly, however, we observe some deviation from the established trends in the case of free parameter set $2$, with the loss function values presented in Table \ref{BH_2_Results_Table} suggesting that it is now the GSL-div that produces the best estimates, despite evidence of convergence to local minima.

\begin{table}[h]

\caption{Calibration Results for Free Parameter Set $2$ of the \citet{Brock_Hommes_1998} Model} \label{BH_2_Results_Table}

\centering

\begin{tabularx}{\textwidth}{lXXXXX}
\hline
& $g_2$ & $b_2$ & $g_3$ & $b_3$ & $L(\hat{\bm{\theta}}, \bm{\theta}_{true})$\\
\hline
$\bm{\theta}_{true}$ & $0.9$ & $0.2$ & $0.9$ & $-0.2$ & $0$ \\
GSL-div/PS & $1.0$ & $0.234$ & $1.0$ & $0$ & $0.2473$ \\
GSL-div/KK & $0.9974$ & $0.3285$ & $0.9325$ & $-0.0675$ & $0.2112$ \\
MSM/PS & $1$ & $0.0485$ & $0.6942$ & $-0.0531$ & $0.3112$ \\
MSM/KK & $0.9742$ & $0.226$ & $0.6691$ & $-0.2272$ & $0.2454$ \\
MIC/PS & $1$ & $0.9332$ & $0.5276$ & $0$ & $0.8523$ \\
MIC/KK & $0.0575$ & $0.9219$ & $0.4587$ & $0$ & $1.2106$ \\
BE & $0.7155$ & $0.3555$ & $0.8382$ & $-0.3481$ & $0.2897$ \\
\hline

\end{tabularx}

\end{table}

\subsubsection{Overall Assessment}

From the preceding computational experiments, it should be apparent that the performance of Bayesian estimation far exceeds that of any of the considered SMD methods. In more detail, we find that for every model and parameter set tested, the method of \citet{Grazzini_et_al_2017} produces estimates that are comparable to the true parameter values, while also being the best performing method for all but the second free parameter set of the \citet{Brock_Hommes_1998} model.

SMD methods, on the other hand, are far less consistent. While there are indeed a number of cases where each objective function performs well, we often observe the emergence of cases where the associated estimates are completely different to the true parameter values. We suggest that two main factors contribute to the emergence of these phenomena. Firstly, in some cases, it appears that a given method may be unable to capture important differences between two datasets. For example, recall that the GSL-div is unable to discriminate between series that differ only by an additive constant and MSM performs poorly when regions of the considered datasets are characterised by different regimes. Secondly, we find that the identification of global minima can become problematic, especially in higher dimensions, resulting in estimates that largely depend on the idiosyncrasies of the considered heuristics.

Given its clear dominance, Bayesian estimation will therefore be the method of choice when attempting to tackle the more ambitious problem of calibrating large-scale ABMs.

\subsection{INET Oxford Housing Market Model}

Although we would ultimately wish to fully calibrate the housing market model, we limit this investigation to two more tractable, though still challenging calibration subproblems involving $4$ and $19$ free parameters respectively. These free parameter sets may seem trivially small in the context of a $100$ parameter model; nevertheless, existing attempts aimed at calibrating models of a similar scale are yet to consider more than $8$ free parameters \citep{Barde_VanDerHoog_2017}, making a set of $19$ free parameters a relatively ambitious undertaking. As in the case of the simple models we previously considered, we provide the technical details associated with these experiments in Appendix \ref{Experiment_Details}.

\subsubsection{Free Parameter Set $1$}

While there are potentially many free parameters that could be considered when constructing our first free parameter set, it is often useful to prioritise parameters that may potentially be strong drivers of the overall model dynamics. Since \textit{Market Average Price Decay}, \textit{Sale Epsilon}, \textit{P Investor} and \textit{Min Investor Percentile} are expected to be relatively important parameters, we consider them in our first calibration attempt.

\begin{figure}[H]

\centering

\includegraphics[width=0.77\linewidth]{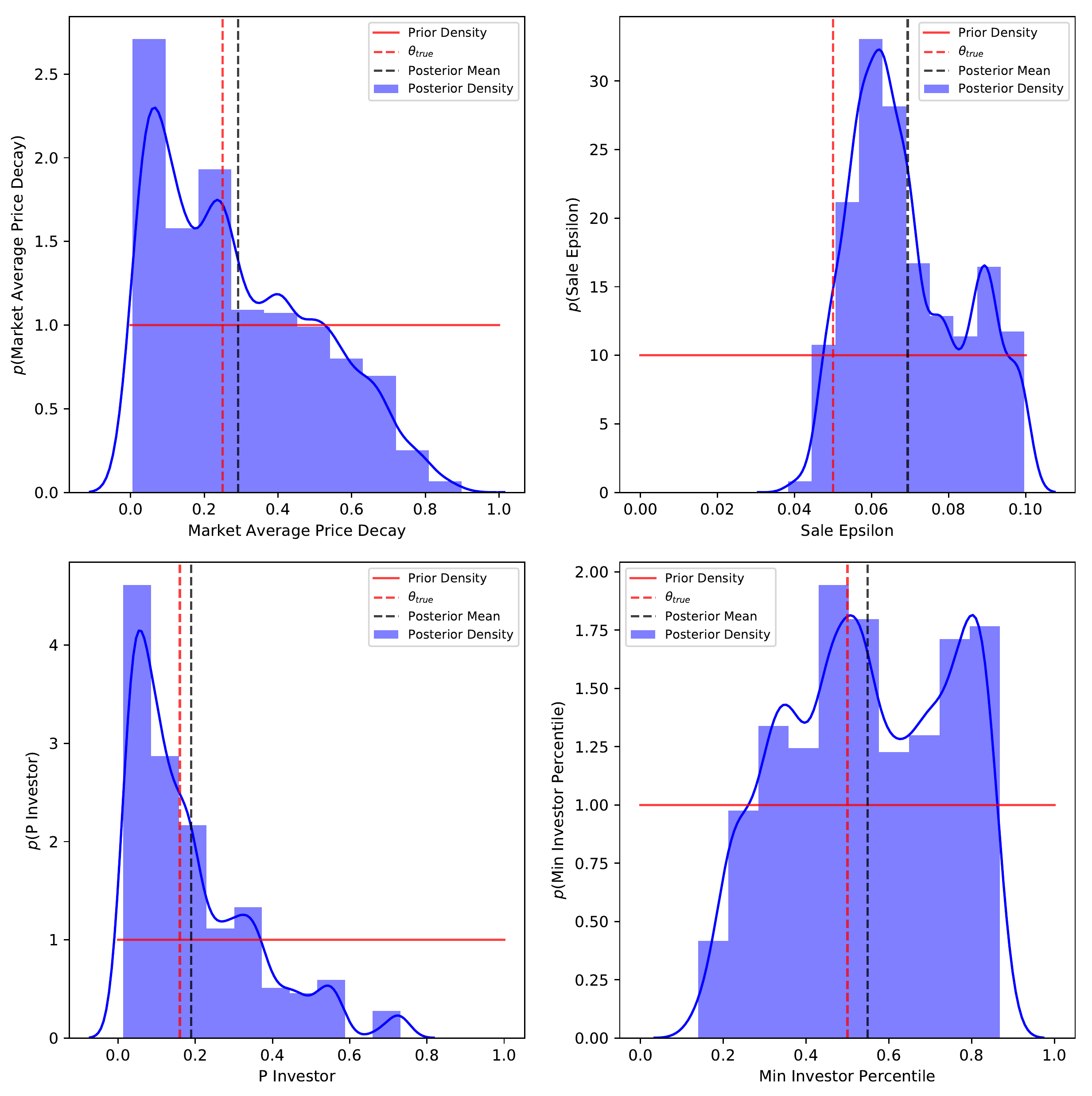}

\caption{Marginal posterior distributions for free parameter set $1$ of the housing market model.} \label{HM_1_Bayesian}

\end{figure}

Referring to Figure \ref{HM_1_Bayesian}, we see that despite a significant increase in the sophistication of the candidate model, the performance of the method of \citet{Grazzini_et_al_2017} is unaffected, resulting in estimates that are reasonably close to the true parameter values and behaviours that are consistent with those observed in the $4$ and $5$ parameter cases of the ARMA$(2, 2)$-ARCH$(2)$ and \citet{Brock_Hommes_1998} models respectively (see Figures \ref{ARMA_ARCH_2_Bayesian} and \ref{BH_2_Bayesian}).

This is a very promising result and suggests that the method is relatively robust and can be trusted in a wide variety of contexts to reliably calibrate several parameters. Nevertheless, we will ultimately need to consider a much larger number of free parameters, which we do shortly, in order to assess the extent to which this performance is maintained as the number of dimensions is increased.

\subsubsection{Free Parameter Set $2$}

\begin{figure}[H]

\centering

\includegraphics[width=0.87\linewidth]{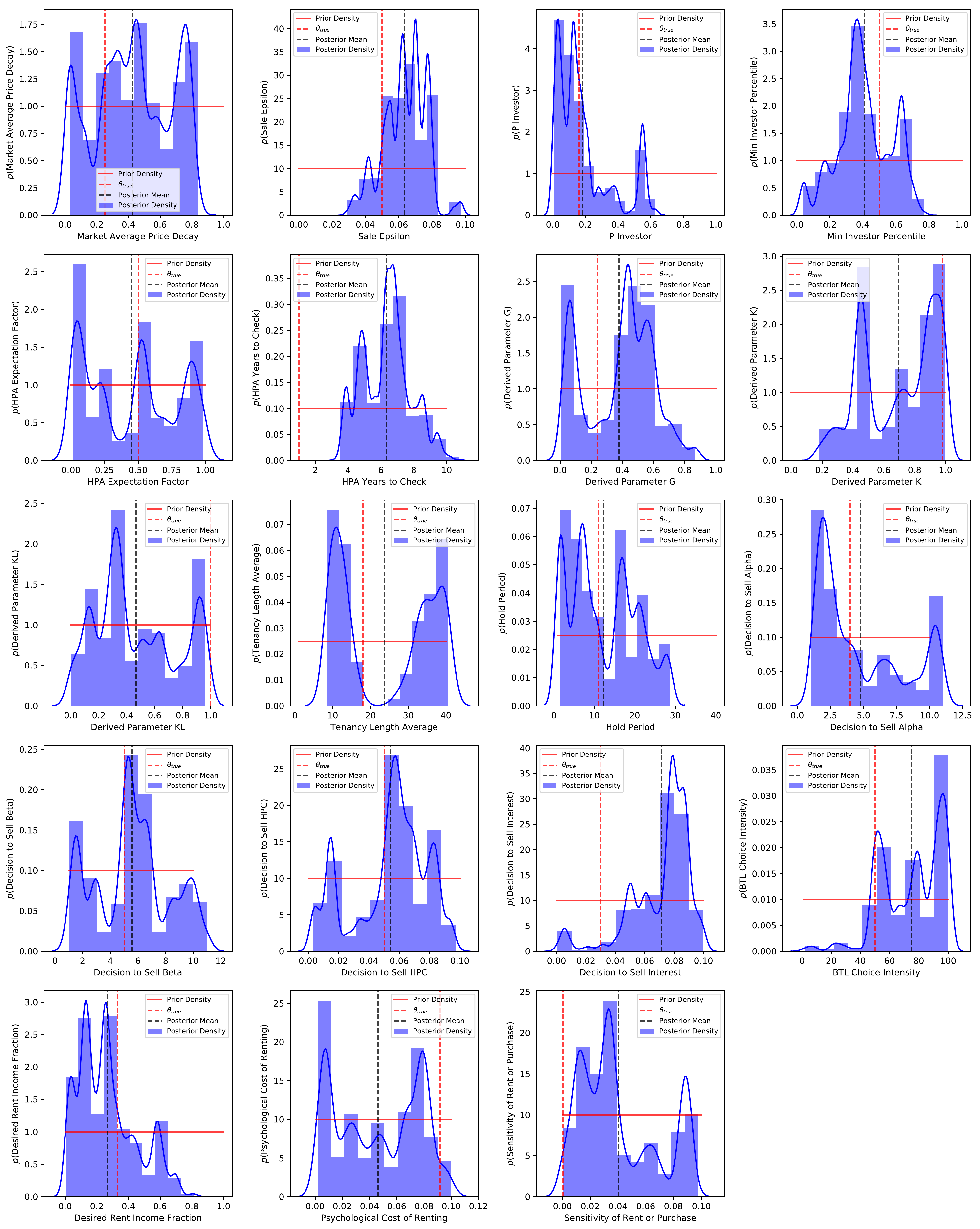}

\caption{Marginal posterior distributions for free parameter set $2$ of the housing market model.} \label{HM_2_Bayesian}

\end{figure}

The second free parameter set, which includes the first as a subset, consists of $19$ parameters that control various aspects of the model's dynamics. We do not list them here for the sake of brevity, though they are clearly indicated in Figure \ref{HM_2_Bayesian}.

Referring to Figure \ref{HM_2_Bayesian}, we observe that:
\begin{enumerate}
\item The estimates obtained for $3$ of the considered free parameters, \textit{P Investor}, \textit{Hold Period}, and \textit{Decision to Sell HPC}, differ from the true parameter values by less than $5\%$ of the explored parameter range. We consider these parameters to have been estimated with a high degree of accuracy.
\item The estimates obtained for $5$ of the considered free parameters, \textit{Min Investor Percentile}, \textit{HPA Expectation Factor}, \textit{Decision to Sell Alpha}, \textit{Decision to Sell Beta}, and \textit{Desired Rent Income Fraction}, differ from the true parameter values by $5-10\%$ of the explored parameter range. Further, the estimates obtained for $3$ of the considered free parameters, \textit{Sale Epsilon}, \textit{Derived Parameter G}, and \textit{Tenancy Length Average}, differ from the true parameter values by $10-15\%$. In both of the aforementioned cases, we consider the parameters in question to have been estimated with a reasonable degree of accuracy.
\item The estimates for the remaining $8$ free parameters differ from the true parameter values by more than $15\%$ of the explored parameter range, with the estimates for $5$ of these parameters differing from the true parameter values by more than $35\%$. In this case, we consider calibration to have been unsuccessful.
\end{enumerate}

We therefore find that the method of \citet{Grazzini_et_al_2017} is able to obtain reasonable estimates for $11$ of the $19$ considered free parameters\footnote{It should also be noted that the obtained posterior distributions demonstrate a far stronger tendency to be multimodal when compared to our previously considered cases, which can affect the accuracy of point estimates \citep{Murphy_2012}.}. While this may seem somewhat disappointing initially, one should bear in mind that, as previously stated, previous investigations involving models of a similar scale have only succeeded in the calibration of $8$ free parameters. Additionally, one should recall that we have made use of only a single time series\footnote{See Appendix \ref{Experiment_Details}.}, rather than the full set of model outputs, and that the method makes very strong independence assumptions. Therefore, it is entirely reasonable to assume that if additional model outputs are considered and the method's assumptions relaxed, Bayesian estimation could provide a robust framework for the calibration of large-scale economic ABMs.

\subsubsection{Goodness of Fit Tests}

\begin{figure}[H]

\centering

\begin{subfigure}{0.48\textwidth}
	\centering
	\includegraphics[width=1\linewidth]{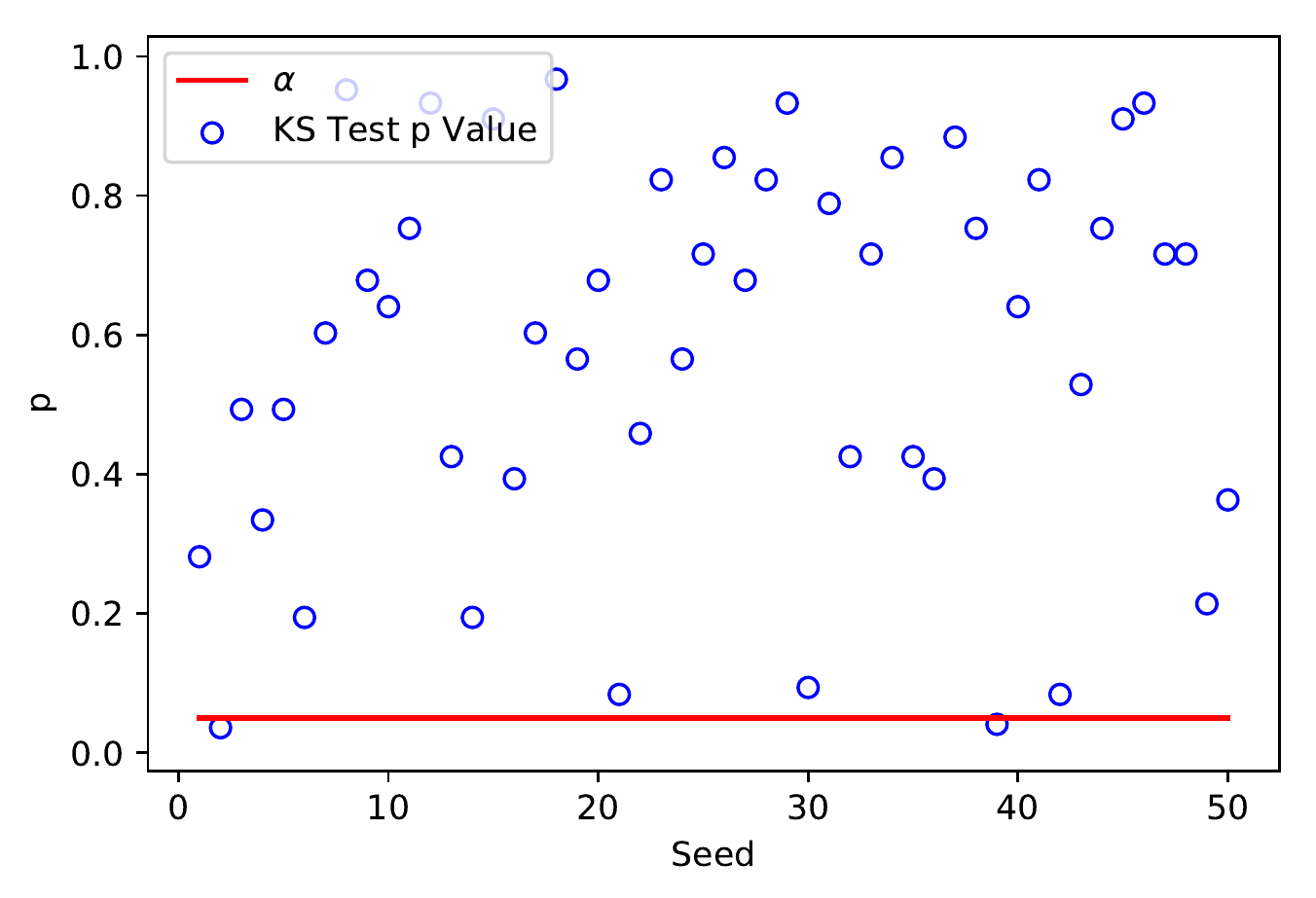}
	\caption{Parameter set $1$.} \label{HM_1_KS}
\end{subfigure}%
\begin{subfigure}{0.48\textwidth}
	\centering
	\includegraphics[width=1\linewidth]{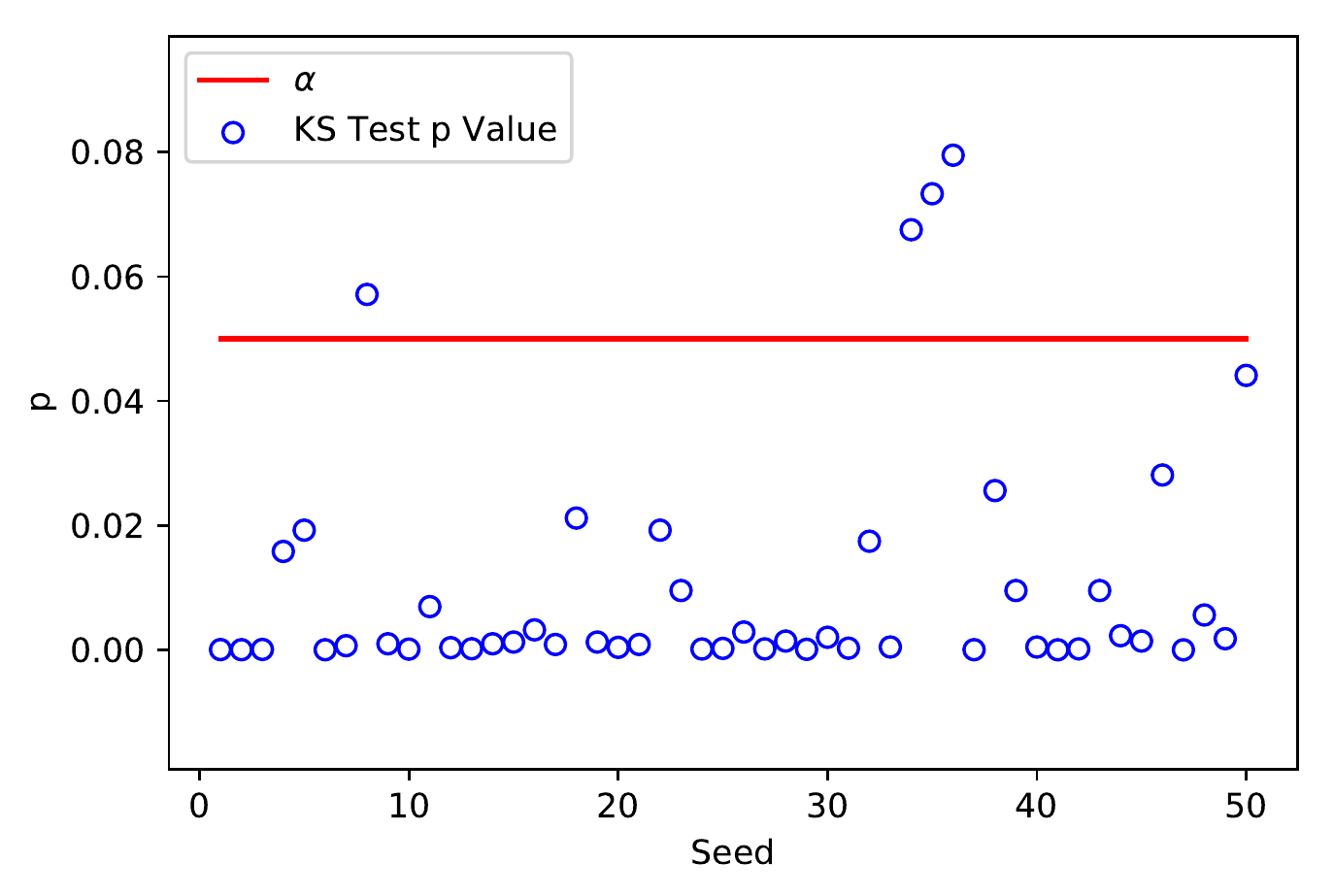}
	\caption{Parameter set $2$.} \label{HM_2_KS}
\end{subfigure}

\caption{Goodness of fit testing for the housing market model.} 

\end{figure}

While defining a comprehensive notion of goodness of fit in the context of ABMs is non-trivial, recall that the method of \citet{Grazzini_et_al_2017} essentially sets parameters such that the unconditional distribution of the model-simulated series matches that of its empirically-sampled equivalent, a notion of goodness of fit that is relatively easy to test using the two-sample Kolmogorov-Smirnov (KS) test. Figure \ref{HM_1_KS} therefore presents the KS test $p$ values obtained when our artificial data is compared to each of the $50$ series in the ensemble of Monte Carlo replications that results when the model is initialised using the parameter values obtained for the first free parameter set using Bayesian estimation. We observe that in all but a small fraction of cases, the KS test suggests that the model-simulated and (artificial) empirically-observed series come from the same distribution, as we would expect, given the proximity of our estimate to the true parameter values.

Repeating the goodness of fit tests for the second free parameter set, we find that the obtained fit is significantly worse, with the KS test suggesting that the model-simulated and (artificial) empirically-observed series come from the same distribution in only a minority of cases (See Figure \ref{HM_2_KS}). This is entirely expected, however, since we obtained inaccurate estimates for $8$ of the considered free parameters, and provides further motivation for efforts to improve existing Bayesian methods.

\section{Conclusions and Future Work}

From the preceding discussions, it should be apparent that despite the impressive progress that has been made during the preceding decade, ABM calibration techniques still require further development before truly robust and general methods can be devised.

In particular, we find that SMD methods, which are favoured in most of the contemporary literature, deliver less compelling performance than is currently suggested by most surveys. This is likely due to the fact that computational experiments of the type we consider here, in which we attempt to recover known parameter values though calibration and compare a variety of methods in identical situations, are almost nonexistent. Furthermore, most existing ABM calibration studies do not emphasise assessments of the uncertainty or accuracy of the associated estimates.

This appears to have led to a lack of awareness regarding the weaknesses of SMD methods and an overall assessment that is perhaps excessively optimistic. As previously suggested, there certainly is some awareness regarding the difficulty associated with minimising SMD objective functions as a result of the haphazard nature of their hypersurfaces and the computational cost of their evaluation. Nevertheless, our computational experiments have revealed that even if these objective functions can be minimised, the associated estimators may suffer from significant a bias in certain situations, which, at this stage, appear impossible to characterise a-priori.

On the other hand, Bayesian methods, which seem to have been largely ignored, deliver far more compelling performance in a wide range of circumstances, failing only when confronted with a large-scale model and a free parameter set of significant cardinality. This would suggest that a paradigm shift is required, with Bayesian methods and the improvement of existing Bayesian estimation techniques needing to become key areas of focus in future research.

Along these lines, we suggest that attempts be made to improve the method of \citet{Grazzini_et_al_2017}, with a particular focus on relaxing the required independence assumptions and allowing for temporal dependencies to be captured when comparing datasets. 

\section*{Acknowledgements}

The author would like to thank Amazon Web Services for the generous provision of access to cloud computing resources and the UK government for the award of a Commonwealth Scholarship. Responsibility for the conclusions herein lies entirely with the author. 


\renewcommand{\refname}{\spacedlowsmallcaps{References}}

\bibliographystyle{abbrvnat}

\bibliography{DPhil_Bibliography}


\appendix

\section{Technical Details \label{Experiment_Details}}

The primary purpose of this appendix is to provide readers with a complete specification of the technical details associated with each computational experiment described in the preceding sections. It is hoped that this will, in principle, allow others to reproduce the presented results.

\subsection{Simple Time Series Models}

\subsubsection{Dataset Construction}

Recall that we do not make use of empirically-observed data. Rather, we consider data generated by a particular model for a chosen set of parameter values and assess the extent to which the true parameter set can be recovered through the calibration of the model to the artificial data. We therefore begin our series of numerical experiments by generating a single dataset for each of the models described in Section \ref{Implemented_Models}, which have been initialised using the parameter values presented in Table \ref{True_Parameters_Simple}.

\begin{table}[H]

\caption{True Parameter Values for the Set of Simple Time Series Models \label{True_Parameters_Simple}}

\centering

\begin{tabularx}{\textwidth}{cXX}
\hline
Model & $\bm{\theta}_{true}$ (Symbols) & $\bm{\theta}_{true}$ (Values) \\
\hline
$1$ & $a_1$ & $0.7$\\
$2$ & $a_0, a_1, a_2, b_1, b_2, c_0, c_1, c_2$ & $0, 0.7, 0.1, 0.2, 0.2, 0.25, 0.5, 0.3$ \\
$3$ & $\tau, \sigma_1, \sigma_2, d_1, d_2$ & $700, 0.1, 0.2, 1, 2$ \\
$4$ & $g_1, b_1, g_2, b_2, g_3, b_3, g_4, b_4, r, \beta$ & $0, 0, 0.9, 0.2, 0.9, -0.2, 1.01, 0, 0.01, 1$ \\
\hline
\end{tabularx}

\vspace{0.5cm}

\caption*{The model numbers above correspond to the order in which the models were introduced in Section \ref{Implemented_Models}.}

\end{table}

\begin{figure}[H]

\centering

\begin{subfigure}{0.5\textwidth}
	\centering
	\includegraphics[width=0.95\linewidth]{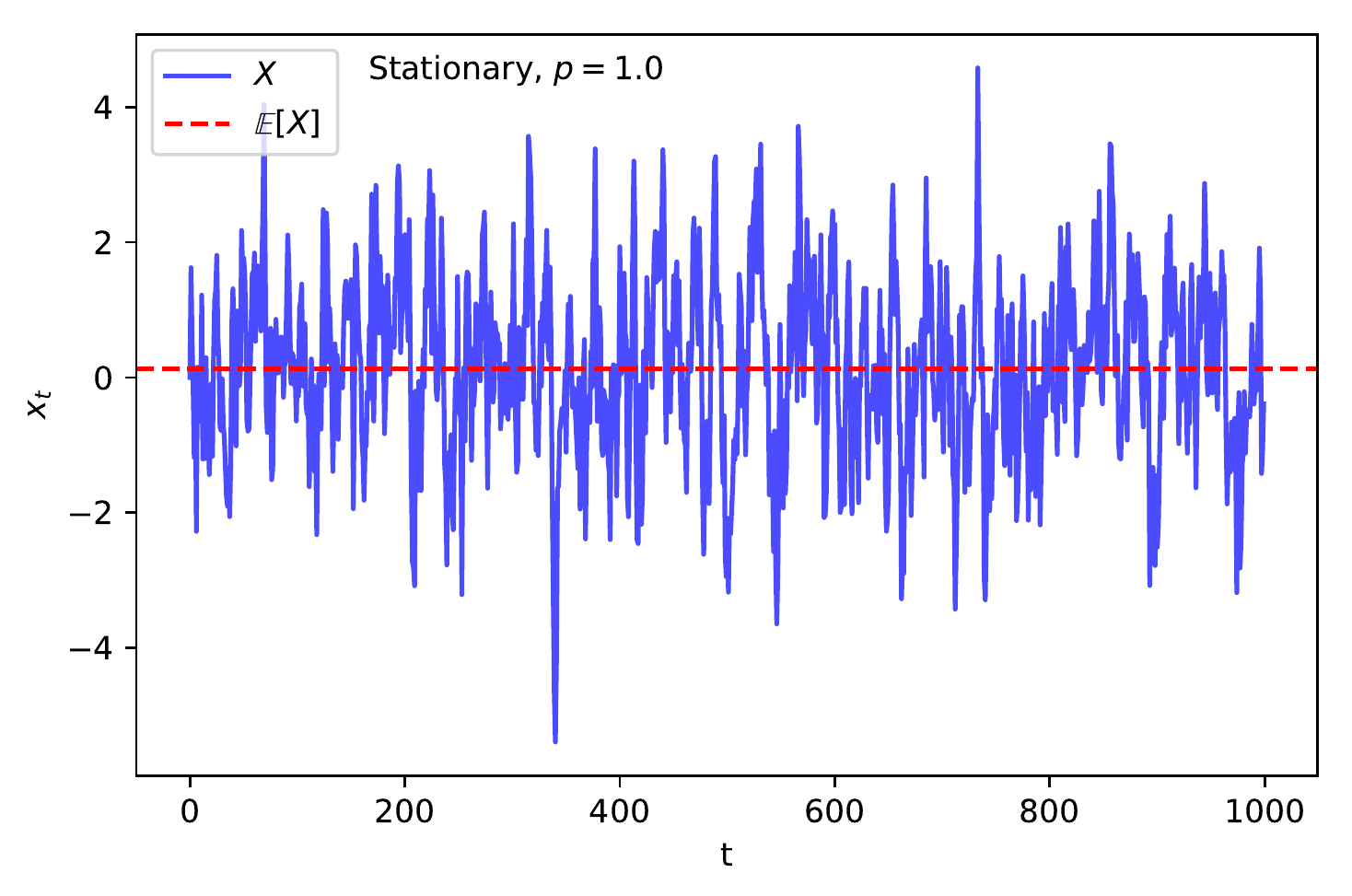}
	\caption{AR($1$)}
\end{subfigure}%
\begin{subfigure}{0.5\textwidth}
\centering
\includegraphics[width=0.95\linewidth]{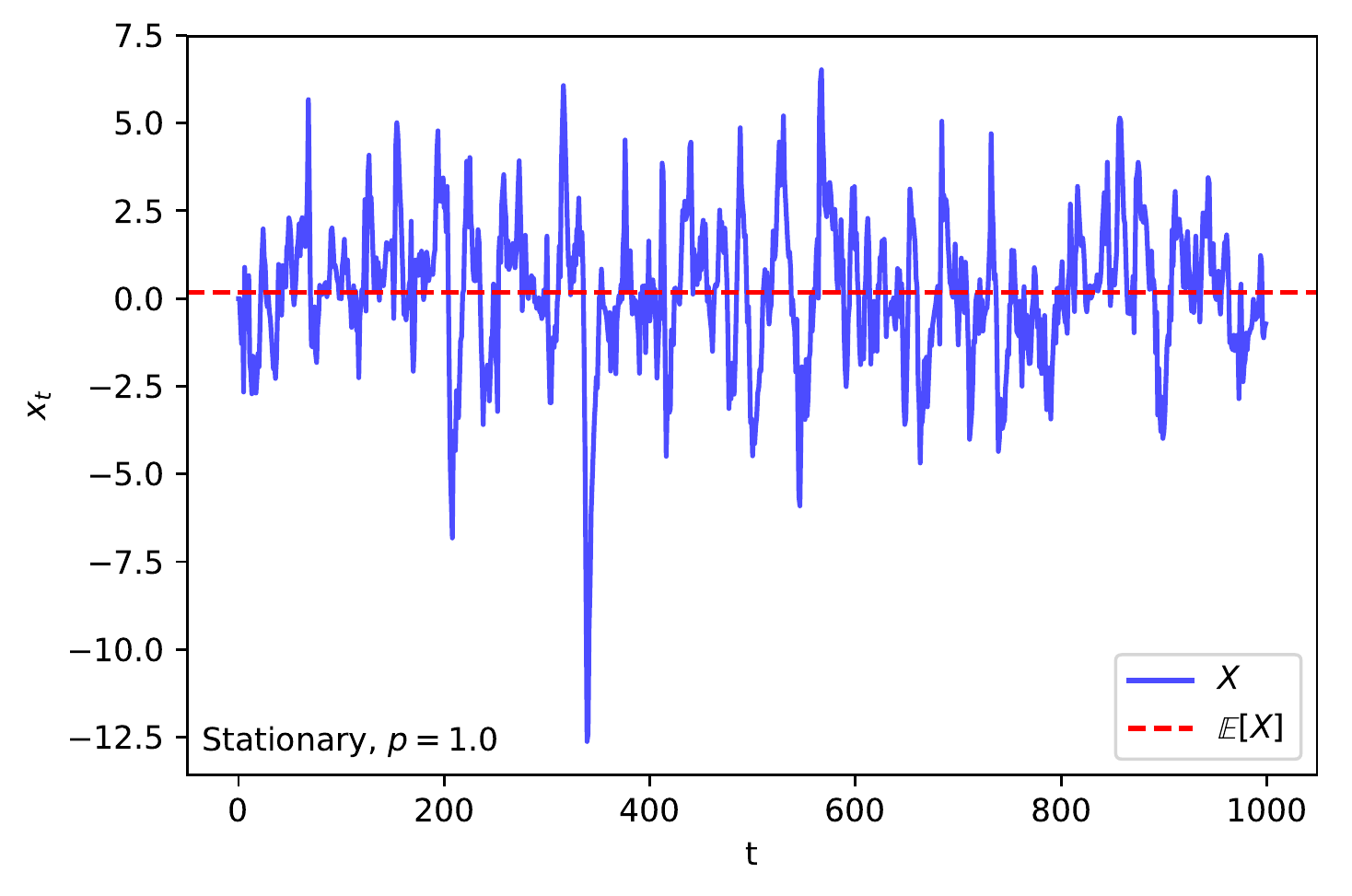}
\caption{ARMA($2, 2$)-ARCH($2$)}
\end{subfigure}

\vspace{0.6cm}

\begin{subfigure}{0.5\textwidth}
	\centering
	\includegraphics[width=0.95\linewidth]{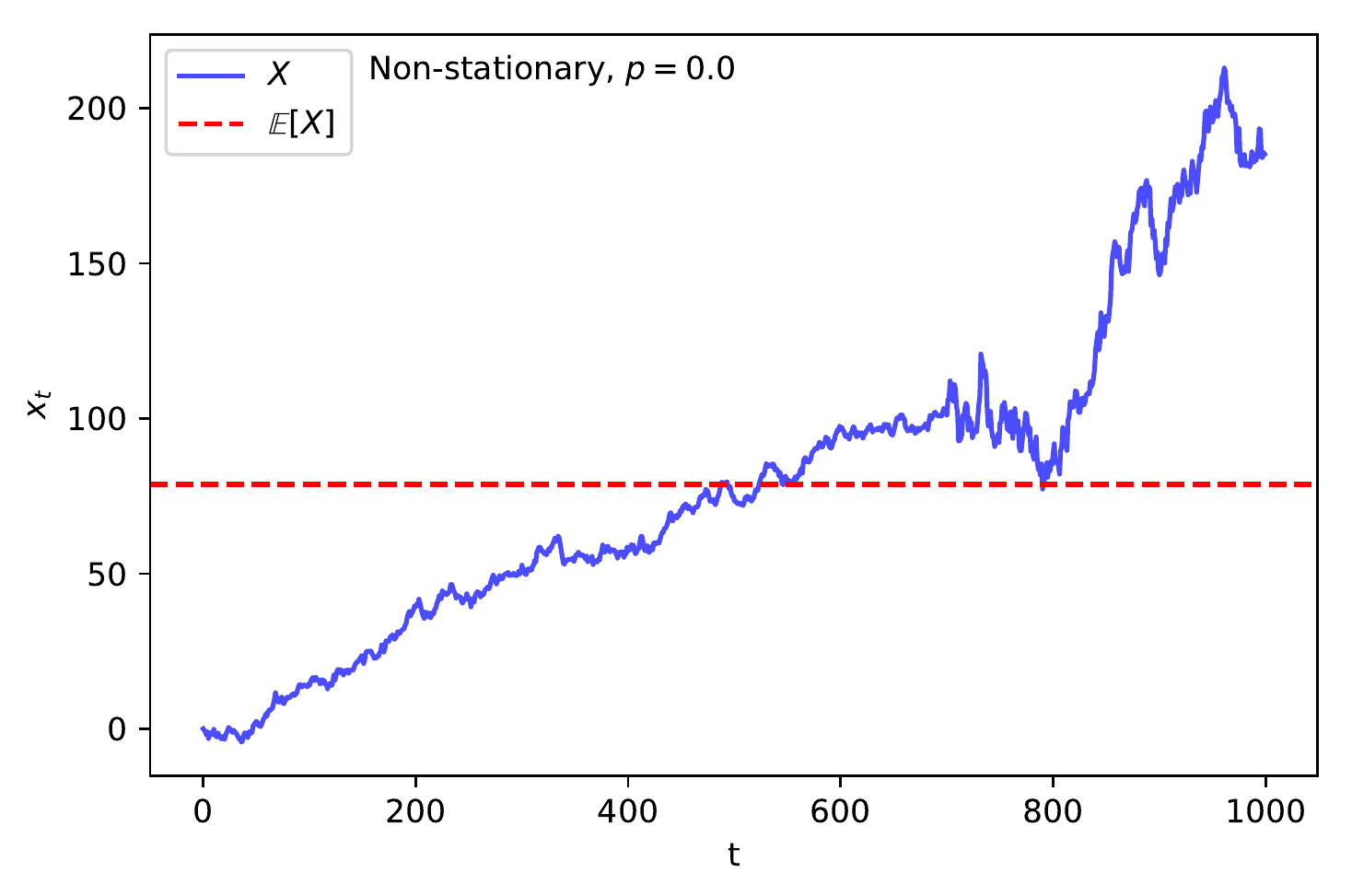}
	\caption{Random walk}
\end{subfigure}%
\begin{subfigure}{0.5\textwidth}
\centering
\includegraphics[width=0.95\linewidth]{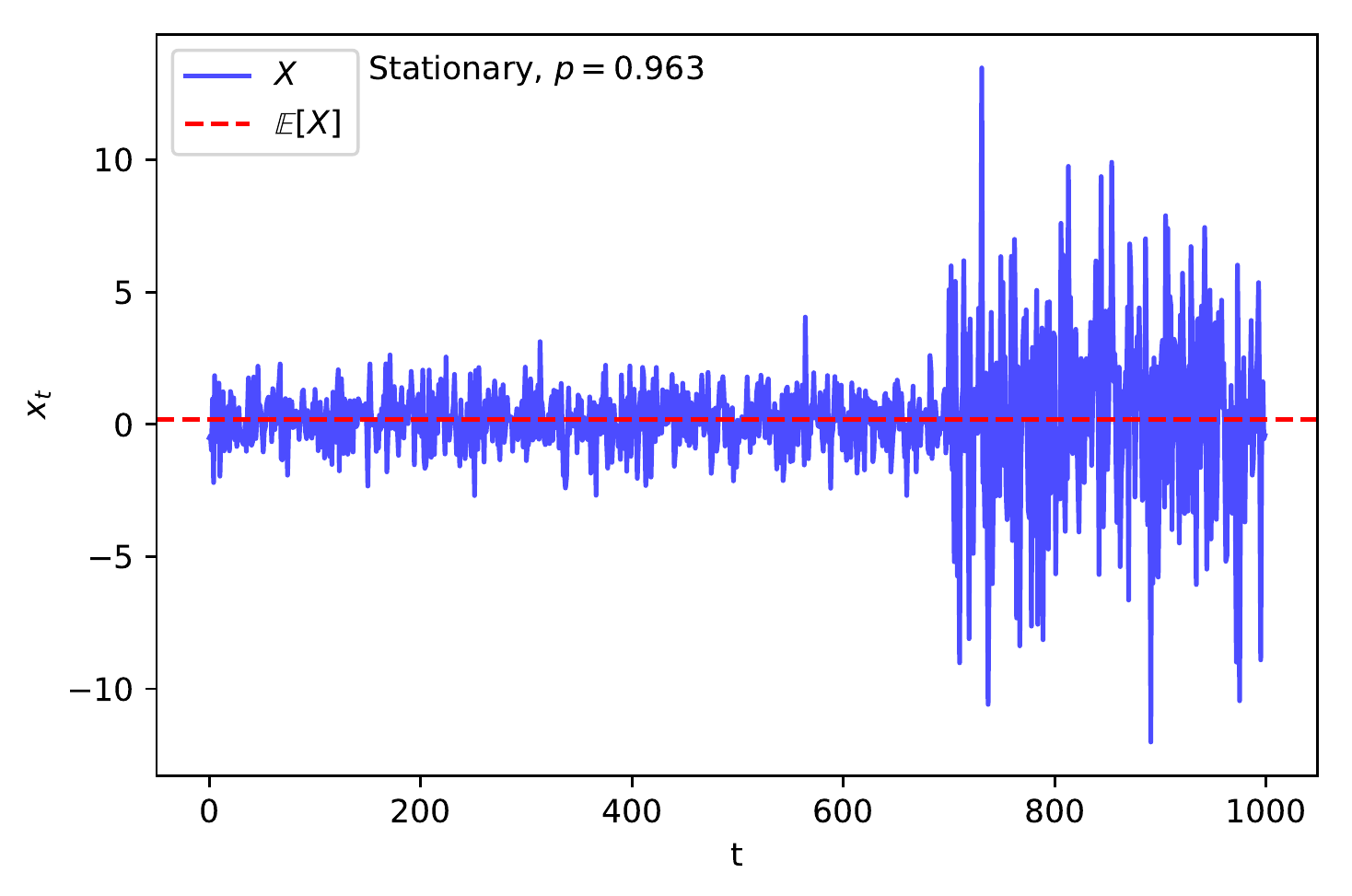}
\caption{Random walk (differences)}
\end{subfigure}

\vspace{0.6cm}

\begin{subfigure}{0.5\textwidth}
\centering
\includegraphics[width=0.95\linewidth]{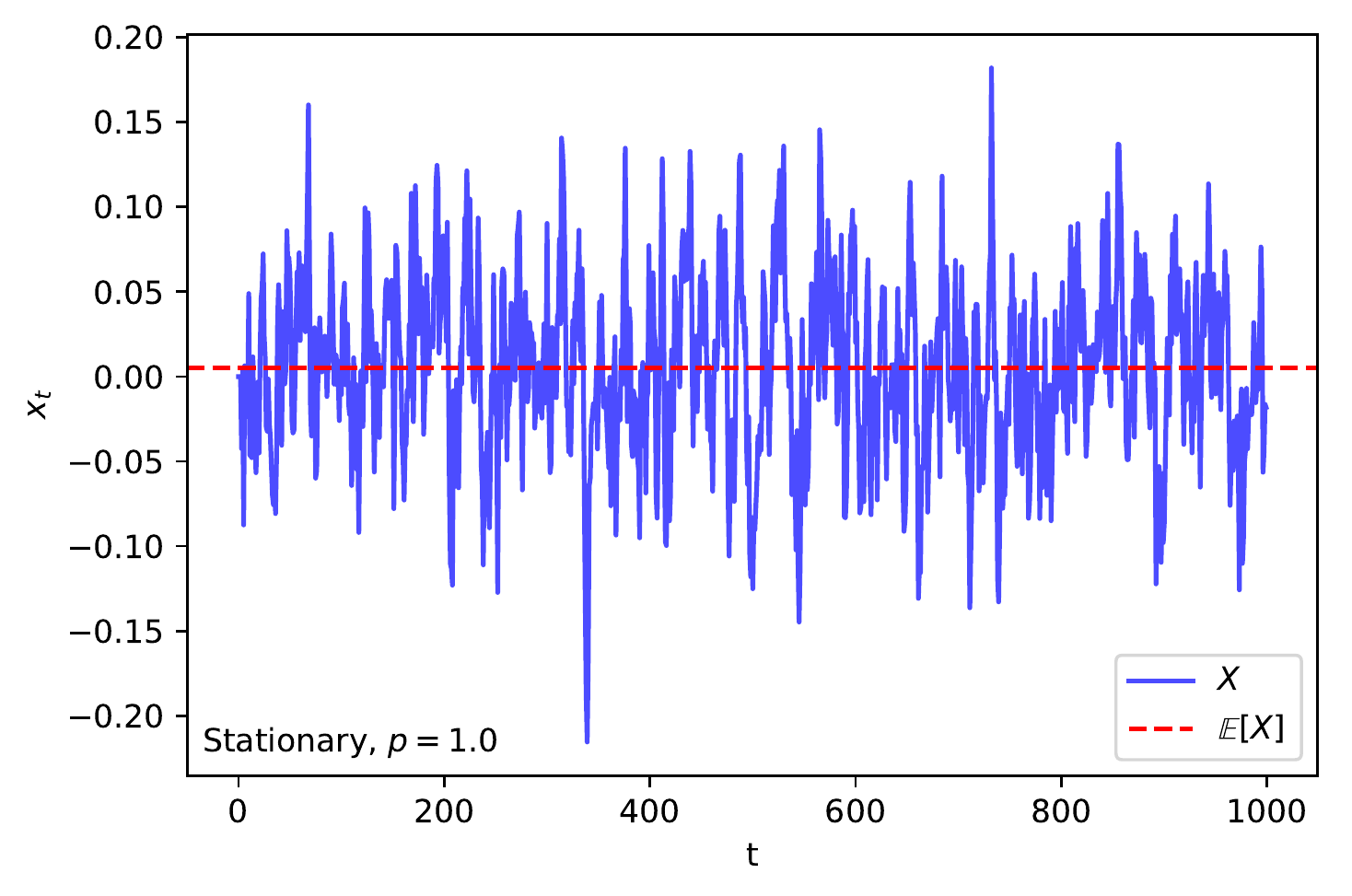}
\caption{\citet{Brock_Hommes_1998}}
\end{subfigure}

\caption{An illustration of the artificial data considered in the calibration experiments. The provided $p$ values are obtained from the stationarity test proposed by \citet{Grazzini_2012}.} \label{Simple_Model_Data}

\end{figure}

It should be noted that each calibration method makes certain assumptions regarding the data and we should therefore ensure that all such assumptions are satisfied by each artificial dataset before proceeding any further. The most notable of these assumptions, and one common to the majority of the considered methods, is that of stationarity. Referring to Figure \ref{Simple_Model_Data}, where we present the artificial data obtained for the selected parameter values, we see that the assumption of stationarity is indeed satisfied in almost all cases, with the only exception being the random walk. This is easily addressed, however, since the output of the random walk can be transformed to induce stationarity by considering the series of first differences\footnote{This transformation is applied to all series generated by the random walk model in all calibration experiments.}.

\subsubsection{Experimental Procedure}

While Section \ref{Experimental_Procedure} provides a relatively complete description of the experimental procedure, it is still necessary to clarify a number of finer details.

Firstly, we are required to choose appropriate lengths for both the empirically-observed and model-simulated time series. Figure \ref{Simple_Model_Data} reveals that we have selected an empirical time series length of $T = 1000$ for all models, which we believe strikes a good balance between being sufficiently long for robust calibration exercises, while still being sufficiently short such that time series data of a similar length could be obtained empirically. We have similarly selected a simulated time series length of $T_{s} = 1000$.

Secondly, we have not assumed ergodicity and must thus consider an ensemble of model-simulated series as opposed to a single realisation in all calibration experiments\footnote{Note that this excludes our artificial datasets. While they are indeed model-generated, we consider them to be equivalent to empirically-observed data.}. In this case, one must be aware of the increased computational cost associated with the simulation of each additional series, while still ensuring that the ensemble is sufficiently large to reduce the variance associated with Monte Carlo simulation to acceptable levels. We therefore consider an ensemble of $250$ realisations in the case of SMD methods\footnote{For series of length $1000$, \citet{Barde_2017} suggests that ensemble sizes of $250$ or $500$ be considered when employing the MIC, motivating our choice.} and $100$ realisations in the case of Bayesian estimation.

At this point, it should be noted that SMD methods generally require a greater number of Monte Carlo replications when compared to the Bayesian estimation approach of \citet{Grazzini_et_al_2017}. This is because most SMD methods involve the estimation of quantities such as moments or occurrence frequencies for each series, which are then averaged over the ensemble. Each additional series therefore provides only one additional value of the quantity of interest. In the case of Bayesian estimation, however, each new series adds an additional $T_s$ observations to the sample to which KDE is applied. 

Finally, we note that while the optimisation algorithms employed in the case of SMD methods can simply be iterated until convergence, we need to specify the desired number of sampled points in the case of the Metropolis-Hastings algorithm used during Bayesian estimation. In this simplified context, we make use of $4$ independent instantiations of the algorithm, each sampling $5000$ points, the first $1500$ of which are discarded. This results in a combined sample consisting of $14000$ sampled points.

\subsubsection{Method Hyperparameters}

It often arises that calibration methods have parameters of their own that need to be carefully chosen in order to maximise performance. In the case of the GSL-div, MSM and Bayesian estimation, we simply adopt the suggestions of the original authors.

\begin{table}[h]

\caption{MIC Hyperparameter Values and Validity Tests \label{MIC_Parameters}}

\centering

\begin{tabularx}{\textwidth}{cllllllllXX}
\hline
Model & & & $b_l$ & $b_u$ & $r$ & $L$ & & & Uniform & Uncorrelated \\
\hline
$1$ & & & $-5$ & $5$ & $5$ & $3$ & & &Yes, $p = 0.5073$ & Yes, $p = 0.0869$ \\
$2$ & & & $-30$ & $30$ & $7$ & $2$ & & & Yes, $p = 0.6858$ & Yes, $p = 0.0989$ \\
$3$ & & & $-15$ & $15$ & $6$ & $3$ & & & Yes, $p = 0.2983$ & Yes, $p = 0.1423$ \\
$4$ & & & $-1$ & $1$ & $8$ & $2$ & & & Yes, $p = 0.9447$ & Yes, $p = 0.6757$ \\
\hline
\end{tabularx}

\vspace{0.5cm}

\caption*{The model numbers above correspond to the order in which the models were introduced in Section \ref{Implemented_Models}.}

\end{table}

The setting of the MIC's hyperparameters is, however, slightly more nuanced, since suitable parameter settings vary from model to model. Fortunately, \citet{Barde_2017} introduces a set of tests that can be applied to ensure that the chosen values are appropriate. Our selected values and the results of these tests are presented in Table \ref{MIC_Parameters}.

It should be noted that the choice $L = 2$ for two of the above models is necessitated due to memory limitations. Nevertheless, \citet{Barde_2017} finds that such a setting is still reasonable, even if not necessary ideal.

\subsection{INET Oxford Housing Market Model}

\subsubsection{Dataset Construction}

Unlike the models we have previously considered, the housing market model produces panel data as opposed to a single time series. While one would ultimately want to make use of the full spectrum of the model's outputs when attempting to calibrate a significant number of parameters, the method of \citet{Grazzini_et_al_2017} is limited to the univariate case. For this reason, we are required to choose a single time series from the model-generated ensemble. Fortunately, this decision is not difficult, with the housing price index (HPI) being a logical choice in this case.

\begin{figure}[H]

\centering

\begin{subfigure}[t]{0.48\textwidth}
\centering
\includegraphics[width=1\linewidth]{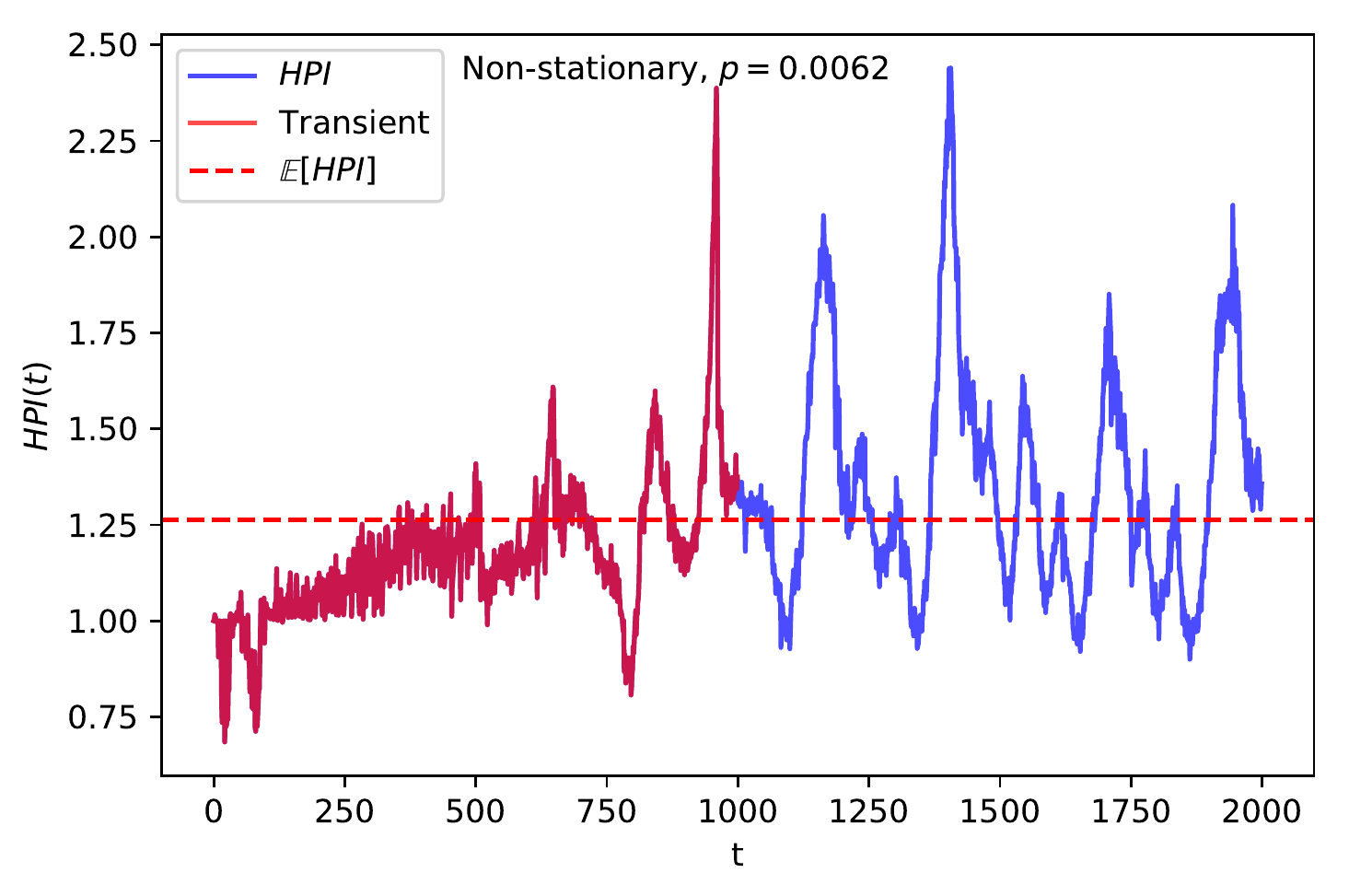}
\caption{Original Time Series} \label{HM_Output}
\end{subfigure}%
\begin{subfigure}[t]{0.48\textwidth}
\centering
\includegraphics[width=1\linewidth]{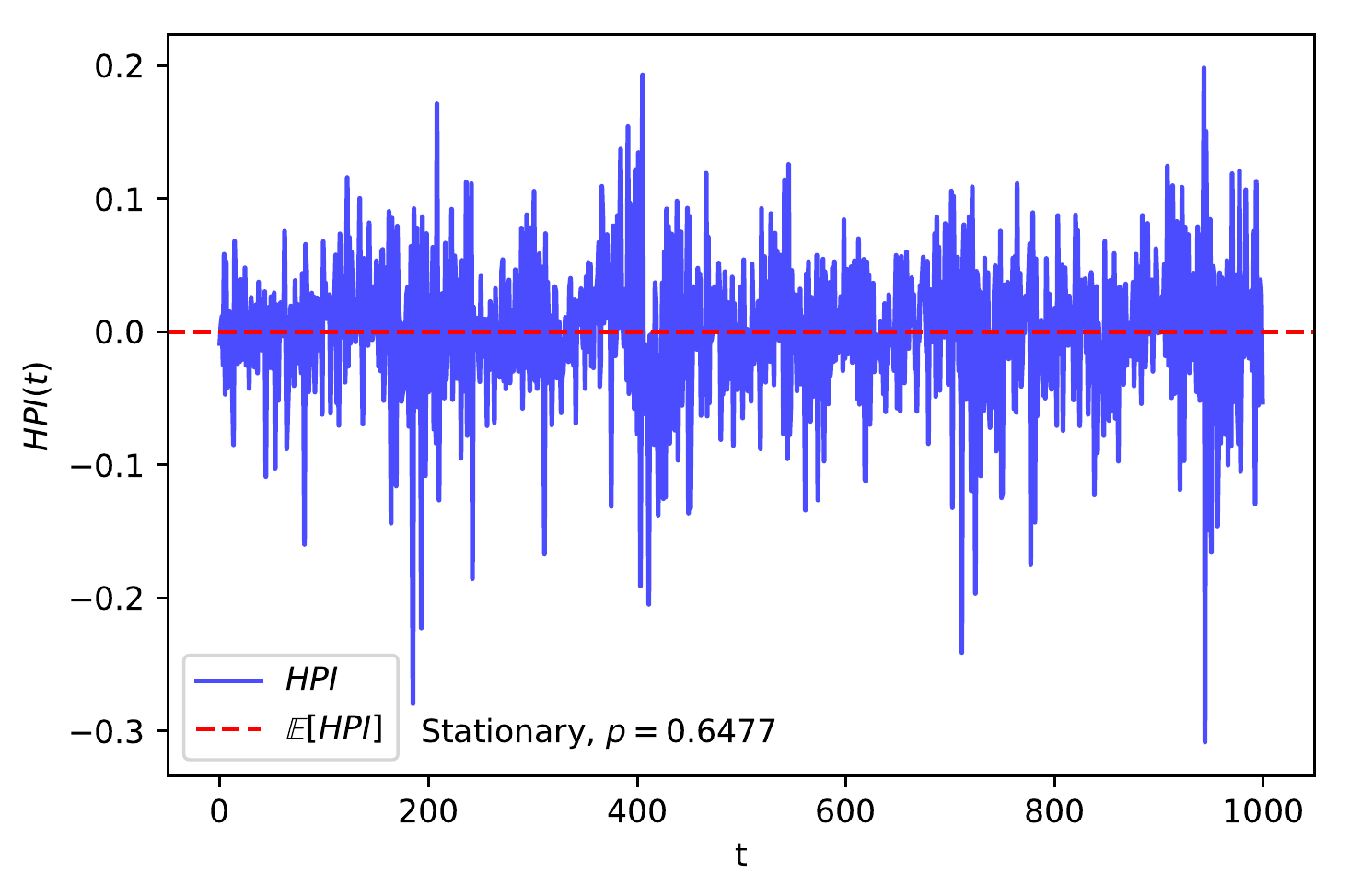} 
\caption{First Difference Series with the Transient Period Discarded} \label{HM_Diff_Output}
\end{subfigure}

\caption{HPI time series for \textit{Market Average Price Decay} $= 0.25$, \textit{Sale Epsilon} $= 0.05$, \textit{P Investor} $= 0.16$, and \textit{Min Investor Percentile} $= 0.5$. The provided $p$ values are obtained from the stationarity test proposed by \citet{Grazzini_2012}.}

\end{figure}

Referring to Figure \ref{HM_Output}, we see that the consideration of HPI results in further complications. Firstly, notice that the HPI time series is characterised by an initial transient period, followed by cyclical dynamics that persist for the remainder of the simulation. Since this transient period corresponds to the model's initialisation procedure rather than its true dynamics, it should be discarded. Secondly, we see that the model output is non-stationary, hence we consider the time series of first differences, as in the case of the random walk model (see Figure \ref{HM_Diff_Output}).

\subsubsection{Experimental Procedure}

Our experimental procedure remains largely unchanged from that employed during the calibration of the set of simple time series models. Nevertheless, the increased computational cost associated with the housing market model requires us to make a number of small adjustments. Firstly, we reduce the number of Monte Carlo replications to $50$, such that the amount of time taken to perform a single evaluation of the posterior density function is more computationally tractable. Secondly, we reduce the overall number of sampled points. In the case of parameter set $1$, we consider $2$ independent instantiations of the Metropolis-Hastings algorithm, each sampling $5000$ points, resulting in a total sample size of $7000$ (after discarding the first $1500$ samples of each instantiation). In the case of parameter set $2$, we again consider $2$ independent instantiations of the Metropolis-Hastings algorithm, though each instantiation now samples $10000$ points, resulting in a total sample size of $17000$ (after burning-in).

\end{document}